\documentclass[12pt]{article}

\usepackage[a4paper,left=30mm,right=20mm,top=20mm,bottom=20mm]{geometry}
\usepackage{graphics}
\usepackage{amsmath}
\usepackage{epsfig}
\usepackage{cite}
\usepackage{xcolor}

\title{Revisiting the production of $J/\psi$ pairs at the LHC}

\author{A.A.~Prokhorov$^{1,2}$, A.V.~Lipatov$^{2,3}$, M.A.~Malyshev$^{3}$, S.P.~Baranov$^4$}

\begin{document}

\maketitle

\begin{center}

{\it $^{1}$Faculty of Physics, Lomonosov Moscow State University, 119991 Moscow, Russia}\\
{\it $^{2}$Joint Institute for Nuclear Research, 141980, Dubna, Moscow region, Russia}\\
{\it $^{3}$Skobeltsyn Institute of Nuclear Physics, Lomonosov Moscow State University, 119991, Moscow, Russia}\\
{\it $^{4}$P.N.~Lebedev Institute of Physics, Moscow 119991, Russia}\\

\end{center}

\vspace{0.5cm}

\begin{center}

{\bf Abstract }

\end{center}

\indent

We consider the prompt double $J/\psi$ production in $pp$ collisions at the LHC
in the framework of $k_T$-factorization QCD approach.
Using the fragmentation mechanism, we evaluate the color octet 
contributions to the production cross sections 
taking into account the combinatorial effects of multiple gluon radiation
in the initial state
driven by the Ciafaloni-Catani-Fiorani-Marchesini 
evolution equation.
We demonstrate the importance of these contributions 
in a certain kinematical region
covered by the CMS and ATLAS measurements.
On the other hand,
the experimental data taken by the LHCb Collaboration
at forward rapidities and 
moderate transverse momenta
can be described well by ${\cal O}(\alpha_s^4)$ color singlet 
terms and contributions from the double parton scattering mechanism.
The extracted value of the effective cross section $\sigma_{\rm eff} = 17.5 \pm 4.1$~mb 
is compatible
with many other estimations based on different final states.

\vspace{1.0cm}


\newpage

\section{Introduction} \indent

Prompt production of $J/\psi$ meson 
pairs at high energies is a very intriguing subject of studies\cite{1,2,3,4}.
It provides
a unique laboratory to investigate the quarkonia production
mechanisms predicted by the non-relativistic QCD (NRQCD) 
factorization\cite{5,6},
which is a rigorous framework for the description
of heavy quarkonia production or decays.
The NRQCD implies a separation of perturbatively 
calculated short distance cross sections for
the production of a heavy quark pair in an intermediate 
Fock state ${}^{2S+1}L_J^{(a)}$ with spin $S$, 
orbital angular momentum $L$, total angular momentum $J$ and 
color representation $a$ from its subsequent non-perturbative transition 
into a physical quarkonium via soft gluon radiation.
The latter is described by the 
long-distance non-perturbative matrix elements (LDMEs), 
which obey certain hierarchy in powers of the relative heavy quark velocity $v$\cite{5,6}.
At the next-to-leading order (NLO), NRQCD can explain the LHC data on 
the prompt $J/\psi$, $\psi^\prime$ and $\chi_c$ transverse momentum distributions (see, for example,\cite{7,8,9,10,11,12,13,14}).
However, it has a long-standing challenge in the 
$J/\psi$ and $\psi^\prime$ polarization and provides inadequate 
description\cite{15,16,17} of the $\eta_c$ production data\footnote{One possible solution, which,
however, implies certain modification 
of the NRQCD rules, has been proposed\cite{18} (see also\cite{19,20,21}).} (see also discussions\cite{22,23,24}).
Studying the $J/\psi$ meson pair production
can shed light on the puzzling aspects above
since $c\bar c$ bound state formation takes place here twice.

In the last few years, significant progress has been made
in the NRQCD evaluations of prompt double $J/\psi$ 
production. 
The complete leading-order (LO) calculations,
including both the color singlet (CS) and color octet (CO) terms,
were done\cite{25}. The relativistic corrections to the $J/\psi$ pair
production are carried out\cite{26}.
The NLO contributions to the CS mechanism
are known\cite{27} and partial tree-level NLO$^*$ contributions to
the both CS and CO terms were calculated\cite{28}.
The latter were found to be essential for both low and 
large transverse momenta, as compared to the LO results\footnote{At the moment, full NLO NRQCD predictions
for double $J/\psi$ production
are not available yet.}.
However,
being comparable with the LHCb measurements\cite{3,4}, 
all these evaluations  
have sizeble discrepancies with the latest CMS\cite{1} and ATLAS\cite{2} data, 
especially at large 
transverse momentum $p_T(J/\psi, J/\psi)$,
invariant mass $m(J/\psi, J/\psi)$ and rapidity 
separation $\Delta y(J/\psi, J/\psi)$ of the $J/\psi$ pairs.
For example, the CMS data
are underestimated by the NRQCD predictions with a factor of about $10$\cite{25,27}. The difference between the 
theoretical calculations and more recent ATLAS data at large 
$p_T(J/\psi, J/\psi)$ or $m(J/\psi, J/\psi)$
is typically smaller but still essential.
It was argued\cite{25} that new processes or mechanisms are needed to better
describe the LHC data.

\begin{figure}
\begin{center}
\includegraphics{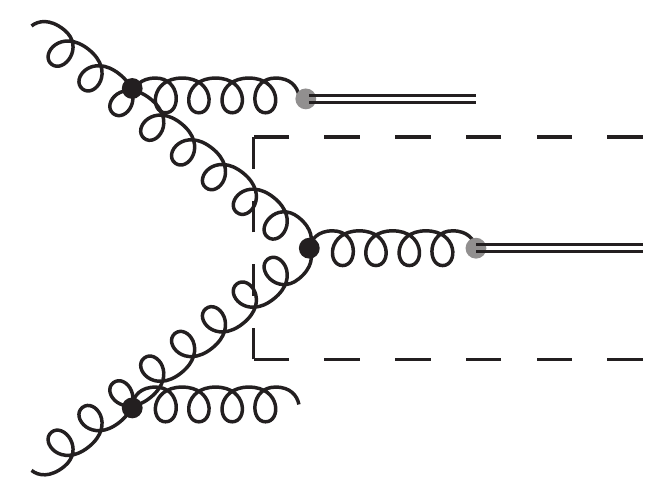}
\caption{Contribution to the $J/\psi$ pair production from the fragmentation of gluon cascade.
The dashed line encloses the hard subprocess $g^* g^*\to g^*$, 
the rest of the diagram describes the initial state radiation cascade.}
\label{fragmgraph}
\end{center}
\end{figure}

At large invariant mass $m(J/\psi, J/\psi)$ the processes with large angular separation
between the $J/\psi$ mesons could play a role.
One of such processes are the gluon or quark fragmentation 
shown in Fig.~\ref{fragmgraph}. The gluon fragmentation into $^3S_{1}^{(8)}$ intermediate state 
scales as $1/p_T^4$
and govern the single $J/\psi$ production at 
high transverse momenta (see, for example,\cite{7,8,9,10,11} and references therein).
In the case of $J/\psi$ pair production,
such terms were found to be negligible since they 
are suppressed by powers of QCD coupling 
$\alpha_s$\cite{25}.
However, at large $p_T(J/\psi, J/\psi)$ or $m(J/\psi, J/\psi)$
one can expect a sizeble combinatorial contribution
to the fragmentation yield
from the multiple gluon radiation originating
during the QCD evolution of the initial gluon cascade.
The latter determines the perturbative QCD corrections to the 
production cross sections at 
high energies, which can be effectively taken into account 
using the Ciafaloni-Catani-Fiorani-Marchesini (CCFM)
evolution equation\cite{29}.
Main goal of our study is to clarify this
point and investigate the role of combinatorial cascade gluon
fragmentation contributions to 
the double $J/\psi$ production
in different kinematical 
regimes at the LHC.

Our other goal is connected
with the investigation of additional 
production mechanism, double parton scattering (DPS), which is
widely discussed in the literature at present (see, for example,\cite{30,31,32,33,34} and references therein).
Apart from the single parton scattering (SPS), where $J/\psi$ meson pair
is produced in a single gluon-gluon collision, DPS events originate from 
two independent parton interactions. 
Studying the DPS mechanism is of great importance since 
it can help in understanding various backgrounds in searches for new 
physics at the collider experiments.
Despite the relative low total production rate, the DPS mechanism is expected to be 
important for double $J/\psi$ production at forward rapidities\cite{35,36,37}.
Therefore, the latter can be used to determine the DPS key parameter, the effective 
cross section $\sigma_{\rm eff}$, which is related to the transverse overlap function between partons in the proton and 
supposed to be universal for all processes with different kinematics and energy scales.
Most of the measured values of $\sigma_{\rm eff}$ lie between $12$ and $20$~mb (see, for example,\cite{38,39}).
However, somewhat lower value $\sigma_{\rm eff} = 8.8 - 12.5$~mb
was extracted from the latest LHCb data on $J/\psi$ pair production
within the NRQCD\cite{4}.
Moreover, the values $\sigma_{\rm eff} = 8.2 \pm 2.2$~mb\cite{40}, $\sigma_{\rm eff} = 6.3 \pm 1.9$~mb\cite{41},
$\sigma_{\rm eff} = 4.8 \pm 2.5$~mb\cite{42} and even $\sigma_{\rm eff} = 2.2 \pm 1.1$~mb\cite{43}, 
$\sigma_{\rm eff} = 2.2 - 6.6$~mb\cite{44} were obtained from 
recent Tevatron and LHC experiments.
Below we will try to extract the effective cross section $\sigma_{\rm eff}$ from  
combined analysis of the LHCb data\cite{3,4} on double $J/\psi$ production taken at $\sqrt s = 7$ and $13$~TeV.

To calculate the physical cross sections we use the $k_T$-factorization 
approach\cite{45,46}. 
We see certain technical advantages in the fact that, even 
with the LO amplitudes for hard subprocesses, one can include
a large piece of higher-order pQCD corrections (NLO + NNLO + ...) taking them into 
account in the form of CCFM-evolved Transverse Momentum Dependent (TMD)
gluon densities in a proton\footnote{The description of this approach 
can be found, for example, in review\cite{47}.}.
In this way we 
preserve consistency with our previous studies\cite{18,19,20,21} and
automatically incorporate the wanted effects of
initial state gluon radiation.
To reconstruct the CCFM evolution ladder,
that is the key point of our consideration,
we employ the TMD parton shower routine implemented into the Monte-Carlo 
event generator \textsc{cascade}\cite{48}.
The $k_T$-factorization approach can be considered as a convenient 
alternative to explicit high-order calculations
in the collinear DGLAP-based scheme. The situation in $J/\psi$ pair
production is specific since calculating even the LO hard scattering amplitudes 
is already complicated enough, so that extending to higher orders
seems to be a rather cumbersome task. Thus, the $k_T$-factorization 
remains the only way open to potentially important higher-order effects.
To evaluate the DPS contributions to the 
double $J/\psi$ production we will use the results of 
our previous analysis\cite{19}.

The outline of the paper is the following. In Section~2 we briefly
describe the basic steps of our calculations. 
In Section~3 we present the numerical results and discussion.
Our conclusions are summarised in Section~4.

\section{The model} \indent

\begin{figure}
\begin{center}
\includegraphics{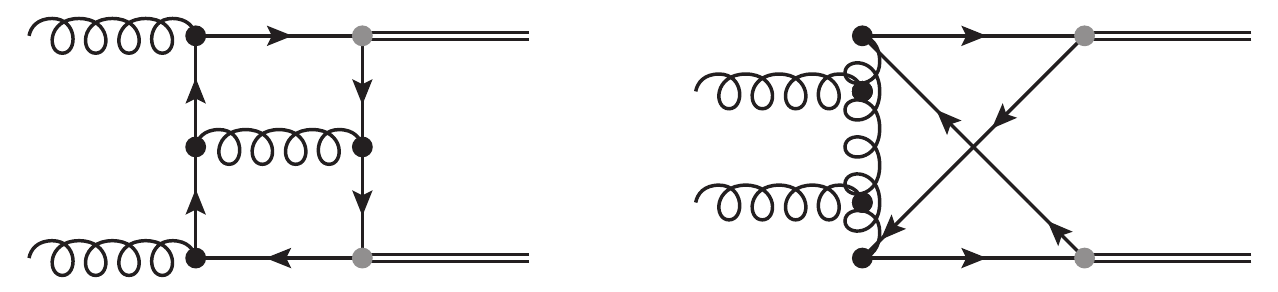}
\caption{Examples of the Feynman diagrams, contributing to the $J/\psi$ pair production via CS mechanism.}
\label{boxgraphs}
\end{center}
\end{figure}

The neccessary starting point of our consideration is related with CS contribution
to the double $J/\psi$ production, that
refers to ${\cal O}(\alpha_s^4)$ gluon-gluon fusion subprocess
\begin{gather}
  g^*(k_1) + g^*(k_2) \to c\bar c\left[^3S_1^{(1)}\right](p_1) + c\bar c\left[^3S_1^{(1)}\right](p_2),
\end{gather}
\noindent
where the four-momenta of all particles are indicated in the parentheses.
Some typical Feynman diagrams are depicted in Fig.~\ref{boxgraphs}.
It is important that both 
initial gluons are off mass shell.
That means that they have
non-zero transverse four-momenta $k_{1T}^2 = - {\mathbf k}_{1T}^2 \neq 0$ and
$k_{2T}^2 = - {\mathbf k}_{2T}^2 \neq 0$ and an admixture of longitudinal 
component in the polarization four-vectors (see\cite{45,46} for more information).
The corresponding off-shell ($k_T$-dependent) production amplitude contains widely used projection 
operators for spin and color\cite{49} 
which guarantee the proper quantum numbers of final state charmonia.
Below we apply the gauge invariant expression obtained earlier\cite{50}.
The derivation steps are explained in detail there.
The respective cross section
can be written as
\begin{gather}
  \sigma(pp \to J/\psi J/\psi + X) = \int {1\over 16 \pi (x_1 x_2 s)^2} |{\cal \bar A}(g^* g^* \to J/\psi J/\psi)|^2 \times \nonumber \\
  \times f_g(x_1,\mathbf k_{1T}^2,\mu^2) f_g(x_2,\mathbf k_{2T}^2,\mu^2) d{\mathbf k}_{1T}^2 d{\mathbf k}_{2T}^2 d{\mathbf p}_{1T}^2 dy_1 dy_2 {d\phi_1\over 2\pi} {d\phi_2\over 2\pi} {d\psi_1\over 2\pi},
\end{gather}
\noindent 
where $\psi_1$ is the azimuthal angle of outgoing $J/\psi$ meson,
$\phi_1$ and $\phi_2$ are the azimuthal angles of  
initial gluons having the longitudinal momentum fractions $x_1$
and $x_2$, $y_1$ and $y_2$ are the center of mass rapidities 
of produced particles and 
$f_g(x,{\mathbf k}_{T}^2, \mu^2)$ is the TMD 
gluon density in a proton taken at the scale $\mu^2$.

In addition to the CS terms above, we have considered some 
of CO contributions using the fragmentation approach.
At high transverse momenta, $p_T \gg m_\psi$,
large logaritmic corrections proportional to 
$\alpha_s^n \ln^n p_T/m_\psi$ occur and, therefore, 
description in terms of fragmentation 
functions (FFs), evolving with the energy scale $\mu^2$, appears to be appropriate.
In general, the FF $D_a^{\cal H}(z,\mu^2)$ 
describing the transition of parton $a$ into the 
charmonium state $\cal H$ 
can be expressed as follows (see, for example,\cite{51} and references therein):
\begin{gather}
  D_a^{\cal H}(z,\mu^2) = \sum_{n}d_a^n(z,\mu^2) \langle {\cal O}^{\cal H} [n]\rangle,
\end{gather} 
\noindent
where $n$ labels the intermediate (CS or CO) state of charmed quark pair
produced in the hard parton interaction
and $\langle {\cal O}^{\cal H} [n]\rangle$ are the corresponding LDMEs.
In the leading logarithmic approximation, 
$g^* \to c\bar c[^3S_1^{(8)}]$ transition is the only one
giving a sizeble contribution 
to $S$-wave charmonia production 
at $p_T \gg m_\psi$\cite{51}, so that the 
cross section of inclusive single $J/\psi$
production in $pp$ collisions could be approximately calculated as
\begin{gather}
  {d\sigma(pp \to J/\psi + X) \over dp_T} = \int dz {d\sigma(pp \to g^*) \over dp_T^{(g^*)}} d_g^{[^3S_1^{(8)}]}(z,\mu^2) \langle {\cal O}^{J/\psi} [^3S_1^{(8)}] \rangle,
\end{gather} 
\noindent
where $p = z p^{(g^*)}$ and $p^{(g^*)}$
are the outgoing $J/\psi$ meson and intermediate gluon momenta. One can easily obtain
\begin{gather}
  \sigma(pp \to g^*) = \int {\pi \over x_1 x_2 s \lambda^{1/2}(m_\psi^2,k_1^2,k_2^2)} |{\cal \bar A}(g^* g^* \to g^*)|^2 \times \nonumber \\ 
  \times f_g(x_1,{\mathbf k}^2_{1T},\mu^2) f_g(x_2,{\mathbf k}^2_{2T},\mu^2) d{\mathbf k}^2_{1T} d{\mathbf k}^2_{2T} dy {d\phi_1 \over 2\pi} {d\phi_2 \over 2\pi},
\end{gather} 
where $p^{(g^*)} = k_1 + k_2$ and $\lambda(m_\psi^2, k_1^2, k_2^2)$ is the known kinematical function\cite{52}.
Evaluation of the off-shell production amplitude 
$|{\cal \bar A}(g^* g^* \to g^*)|^2 = (3/2) \pi \alpha_s(\mu^2) |{\mathbf p}_T^{(g^*)}|^2$
is an extremely straightforward and, in our opinion, needs no explanation.
We only note that, according to the $k_T$-factorization prescription\cite{45,46},
the summation over the polarizations of initial off-shell gluons is carried out with
$\sum \epsilon^\mu \epsilon^{*\,\nu} = {\mathbf k}_T^\mu {\mathbf k}_T^\nu/{\mathbf k}_T^2$.
In the collinear limit ${\mathbf k}_T \to 0$ this expression converges to the ordinary one after averaging on the
azimuthal angle.

The key point of our consideration is
that the gluon, produced in the hard scattering
and fragmented into the $J/\psi$ meson according to main formula~(4), 
is accompanied by a number of gluons radiated during the non-collinear QCD evolution,
which also give rise to final $J/\psi$ mesons.
Thus, taking into account all their possible combinations into 
the meson pairs, one 
can calculate corresponding gluon fragmentation
contribution to the double $J/\psi$
production up to all orders in the pQCD expansion.
At high energies, the QCD evolution of gluon cascade 
can be described by the CCFM equation\cite{29}, which smoothly 
interpolates between the small-$x$ BFKL gluon dynamics 
and high-$x$ DGLAP one, and, therefore, provides us with the suitable tool 
for our phenomenological study.
To reconstruct the CCFM evolution ladder,
we generate a Les Houches Event file\cite{53}
in the numerical calculations according to~(4) and~(5) and then process the file with a 
TMD shower tool implemented into the Monte-Carlo event generator \textsc{cascade}\cite{48}.
This approach gives us the possibility to take into account
the contributions from initial state gluon emissions in a
consistent way (see also\cite{54}).

Of course, the same scenario can be applied to fragmentation
of charmed quark pairs into $J/\psi$ mesons. 
So, one can first simulate the perturbative production of $c \bar c$ pair 
in the off-shell gluon-gluon fusion and then reconstruct the 
CCFM evolution ladder using the \textsc{cascade} tool.
After that, one can easily produce $J/\psi$ pairs by taking into account 
all possible combinations of mesons
originating from the charmed quarks and/or cascade gluon fragmentation. 
Unlike the conventional (collinear) QCD factorization, where only fragmentation of both charmed quarks 
into $J/\psi$ mesons gives contribution, the model above 
can lead to increase in the double $J/\psi$ production 
cross section due to additional combinatorial contributions from
gluons and quarks.

\begin{figure}
\begin{center}
\includegraphics[width=6.0cm]{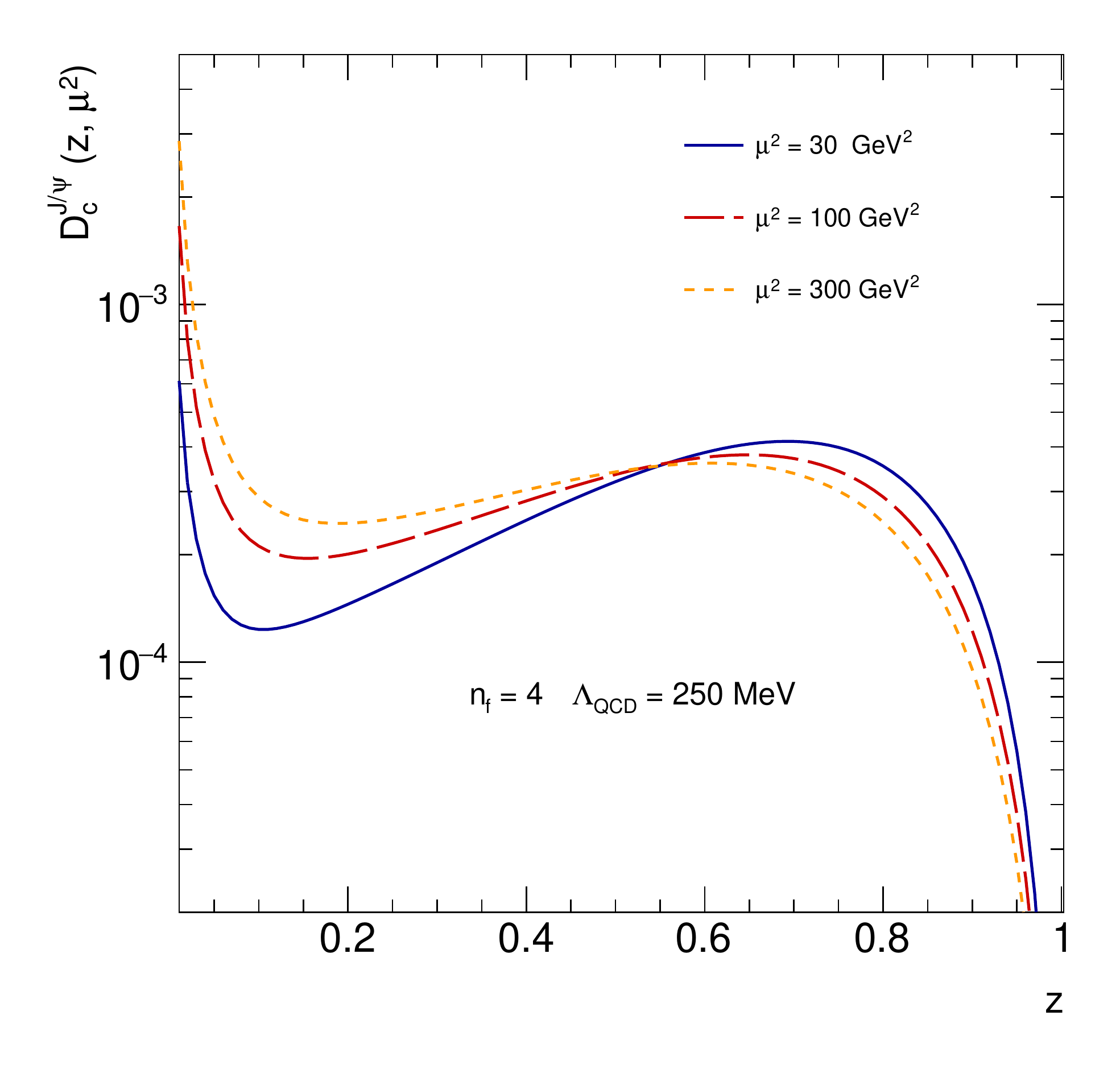}
\includegraphics[width=6.0cm]{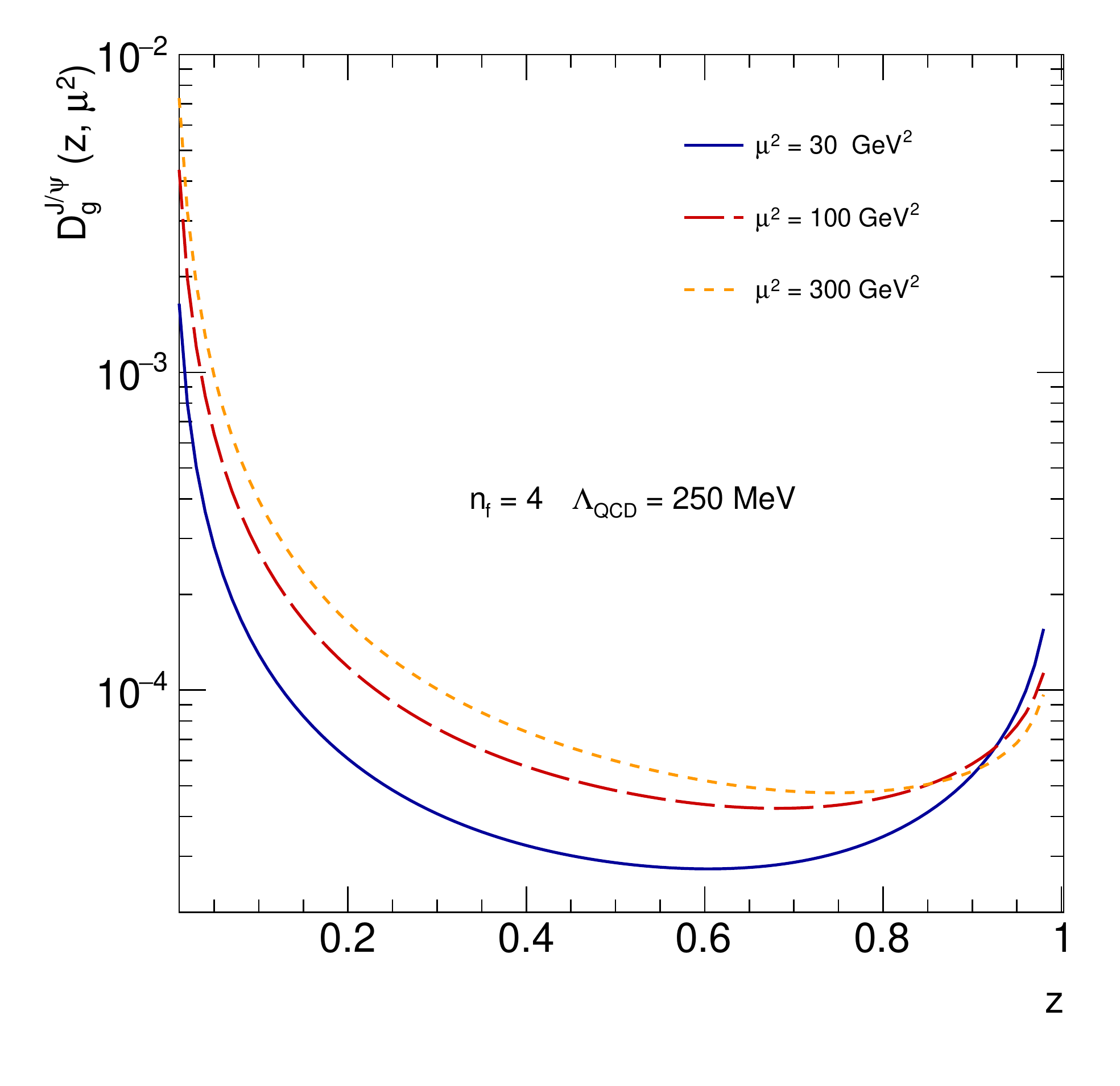}
\caption{The charm (left panel) and gluon (right panel) FFs, $D_c^{J/\psi}(z,\mu^2)$ and $D_g^{J/\psi}(z,\mu^2)$,
calculated as functions of $z$ for several values of scale $\mu^2$. 
We have applied $n_f = 4$, $\Lambda_{\rm QCD} = 250$~MeV,
$\langle {\cal O}^{J/\psi} [^3S_1^{(1)}] \rangle = 1.16$~GeV$^3$ and 
$\langle {\cal O}^{J/\psi} [^3S_1^{(8)}] \rangle = 2.5 \cdot 10^{-3}$~GeV$^3$.}
\label{FFs}
\end{center}
\end{figure}

\begin{figure}
\begin{center}
\includegraphics[width=6.0cm]{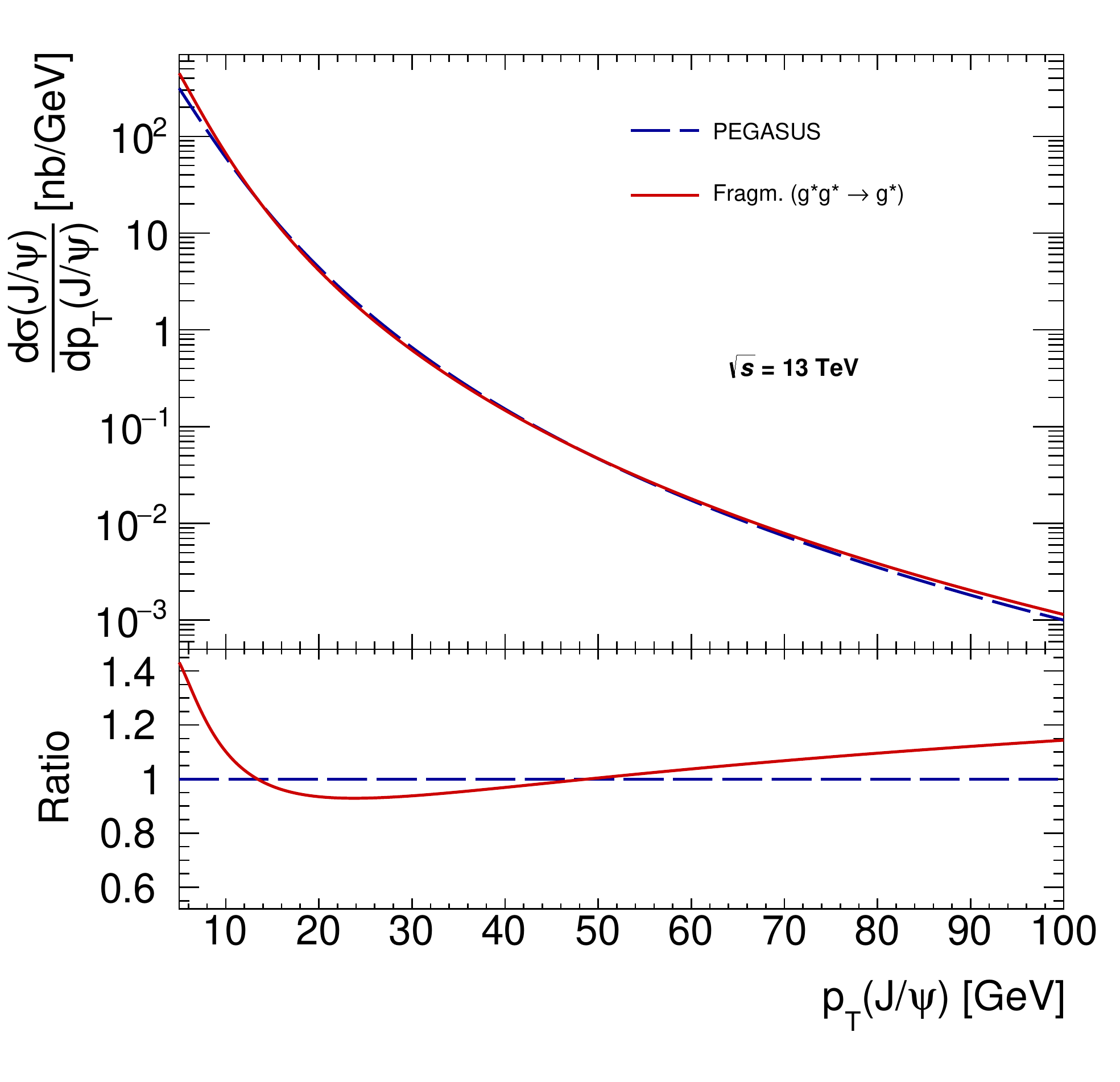}
\includegraphics[width=6.0cm]{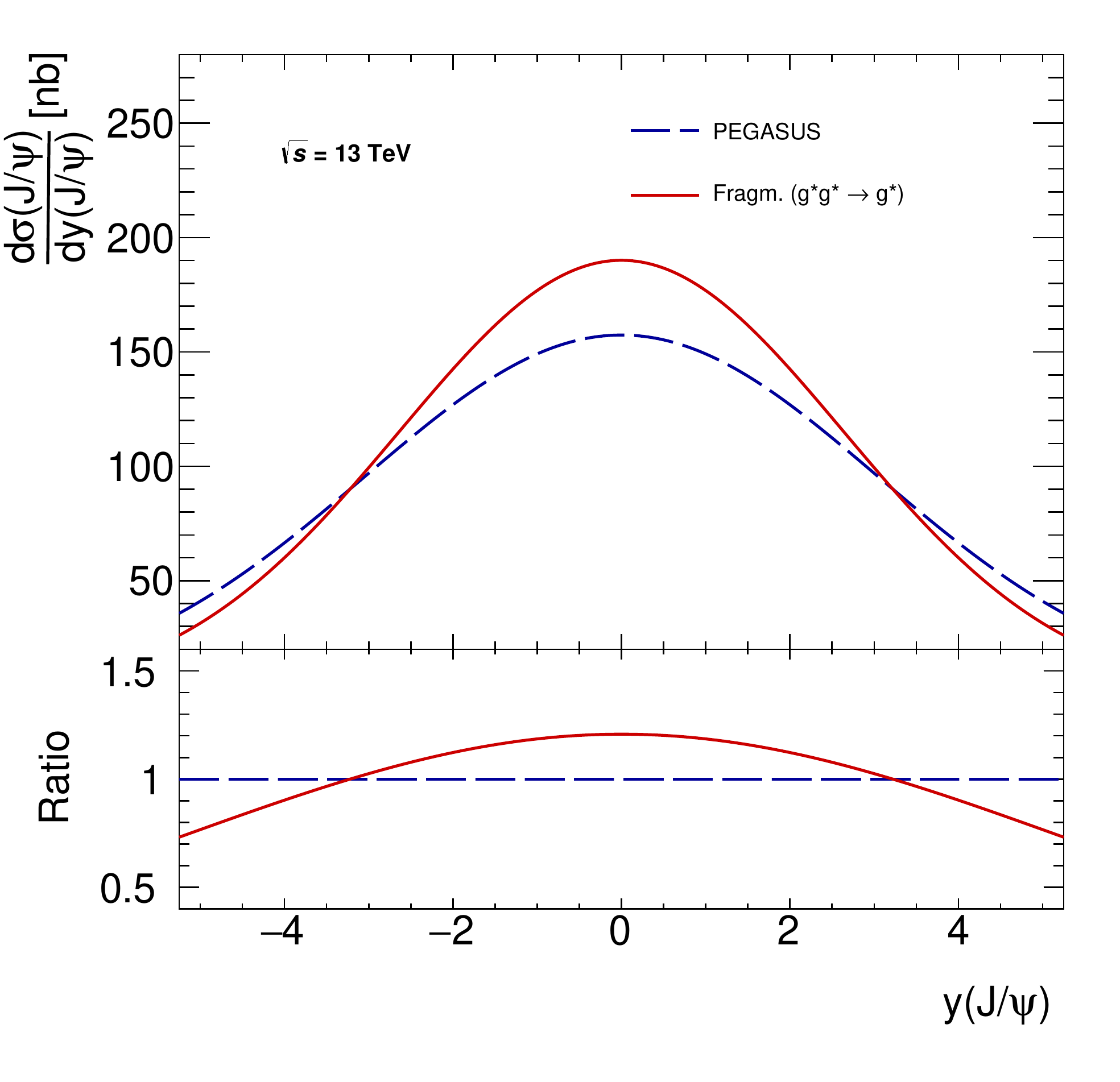}
\caption{Transverse momentum and rapidity distributions
of inclusive $J/\psi$ production at $\sqrt s = 13$~TeV, calculated using the fragmentation approach and 
Monte-Carlo event generator \textsc{pegasus}\cite{55}.
The contributions from the $^3S_1^{(8)}$ transition 
are only taken into account with 
$\langle {\cal O}^{J/\psi} [^3S_1^{(8)}] \rangle = 2.5 \cdot 10^{-3}$~GeV$^3$ \cite{19}.
The A0 gluon distribution in proton is applied.}
\label{PEGASUS_3S1_8}
\end{center}
\end{figure}

The charm and gluon FFs at the any scale $\mu^2$, 
$D_c^{J/\psi}(z,\mu^2)$ and $D_g^{J/\psi}(z,\mu^2)$, 
can be obtained by solving 
the LO DGLAP evolution equations:
\begin{gather}
  \frac{d}{d\ln \mu^2}\left(\begin{array}{lr} D_c \\ D_g \end{array}\right) = \frac{\alpha_s(\mu^2)}{2\pi}\left(\begin{array}{lr} P_{qq} & P_{gq} \\ P_{qg} & P_{gg} \end{array}\right) \otimes \left(\begin{array}{lr} D_c \\ D_g \end{array}\right),
\end{gather} 
\noindent
where $P_{ab}$ are the standard LO DGLAP splitting functions. 
The initial conditions for these FFs are calculated with\cite{51}
\begin{gather}
  d_g^{[^3S_1^{(8)}]}(z,\mu_0^2)= \frac{\alpha_s(\mu_0^2)}{24 m_c^3} \pi \delta(1 - z), 
\end{gather} 
\begin{gather}
  d_c^{[^3S_1^{(1)}]}(z,\mu_0^2)= \frac{\alpha_s^2(\mu_0^2)}{m_c^3} {16z(1-z)^2 \over 243(2-z)^6} \left(5z^4-32z^3+72z^2-32z+16\right),
\end{gather} 
\noindent
where starting scale $\mu_0^2 = m^2_\psi$. 
As it was noted above, we keep only the leading 
contributions to corresponding FFs (see, for example,\cite{51} and references therein).
According to the non-relativistic QCD approximation, we
set the charmed quark mass to $m_c = m_\psi/2$
and then solve the DGLAP equations~(6) numerically.
The obtained charm and gluon FFs, $D_c^{J/\psi}(z,\mu^2)$ and $D_g^{J/\psi}(z,\mu^2)$,
are shown in Fig.~\ref{FFs} as functions of $z$ for several values of scale $\mu^2$. 
Using these FFs, we reproduce well the 
results of calculations performed with the Monte Carlo 
event generator \textsc{pegasus}\cite{55} (see Fig.~\ref{PEGASUS_3S1_8}).

Finally, we turn to the DPS contribution to the double $J/\psi$ production.
We apply a commonly used
factorization formula (for details
see reviews\cite{30,31,32,33,34} and references therein):
\begin{gather}
  \sigma_{\rm DPS}(pp \to J/\psi J/\psi + X) = {1\over 2} {\sigma^2(pp \to J/\psi + X) \over \sigma_{\rm eff}},
\end{gather}
\noindent
where factor $1/2$ accounts for two identical particles in the final state. The 
effective cross section $\sigma_{\rm eff}$ can be considered as 
a normalization constant which incorporates all "DPS unknowns" 
in to a single phenomenological parameter.
Derivation of the factorization formula~(9) relies on the two approximations: first, 
the double parton distribution function can be decomposed into 
longitudinal and transverse components and, second, the longitudinal 
component reduces to the diagonal product of two independent single parton 
densities. The latter is generally acceptable for such collider experiments 
where small-$x$ values are probed.
The typical values of the variable $x$
in the considered process are of order
$x \sim (2m_\psi^2 + p_T^2)^{1/2}/\sqrt s \sim 10^{-3}$,
that approximately corresponds to the kinematical region
of CMS\cite{1}, ATLAS\cite{2} and even LHCb\cite{3,4} measurements
(due to relatively small invariant mass of produced $J/\psi$ pair, see discussion below).
Therefore, one can safely omit the kinematical constraint\cite{56,57} often applied 
at the edge of phase space\footnote{Phenomenological consequences of the kinematical constraint\cite{56,57}
at the large $x$ were investigated\cite{58}.}.
Detailed description of evaluation of inclusive cross section
$\sigma(pp \to J/\psi + X)$ in the $k_T$-factorization approach supplemented with the 
NRQCD formalism can be found\cite{19}.

In the numerical calculations below we will use TMD gluon density in a proton
obtained\cite{59} from the numerical solution of CCFM evolution equation (namely, A0 set), 
where the input parameters have been fitted to the proton structure function $F_2(x,Q^2)$.
At present, the A0 gluon distribution function is widely used in the phenomenological 
applications\footnote{A comprehensive collection of the TMD gluon 
densities can be found in the \textsc{tmdlib} package\cite{60},
which is a C++ library providing a framework and an interface to the different parametrizations.}
(see, for example,\cite{19,20,21}).
The renormalization and factorization scales, $\mu_R$ and $\mu_F$,
were set to $\mu_R^2 = \mu_F^2 = \hat s + {\mathbf Q}^2_T$, where $\hat s = (k_1 + k_2)^2$ and ${\mathbf Q}_T^2$
is the transverse momentum of initial off-shell gluon pair. 
This choice is dictated mainly by the CCFM 
evolution algorithm (see\cite{59} for more information).
As it is often done, 
the fragmentation scale $\mu_{\rm fr}$ is choosen to be equal to $\mu_{\rm fr} = m_T$, the 
transverse mass of fragmenting parton.
We use the one-loop formula for the QCD coupling $\alpha_s$
with $n_f = 4$ active quark flavors at $\Lambda_{\rm QCD}^{(4)} = 250$~MeV.
Following\cite{61}, we set the $J/\psi$ meson mass $m_\psi = 3.097$~GeV.
We take corresponding CS LDME from the known
$J/\psi \to \mu^+\mu^-$ decay width: $\langle {\cal O}^{J/\psi} [^3S_1^{(1)}]\rangle = 1.16$~GeV$^3$\cite{7,8,9,10,11}.


\section{Numerical results and discussion} \indent

We are now in a position to present the results of our simulations.
First we discuss the role of cascade gluon fragmentation
in different kinematical regimes, which correspond to 
the CMS, ATLAS and LHCb experiments.

\begin{figure}
\begin{center}
\includegraphics[width=6.0cm]{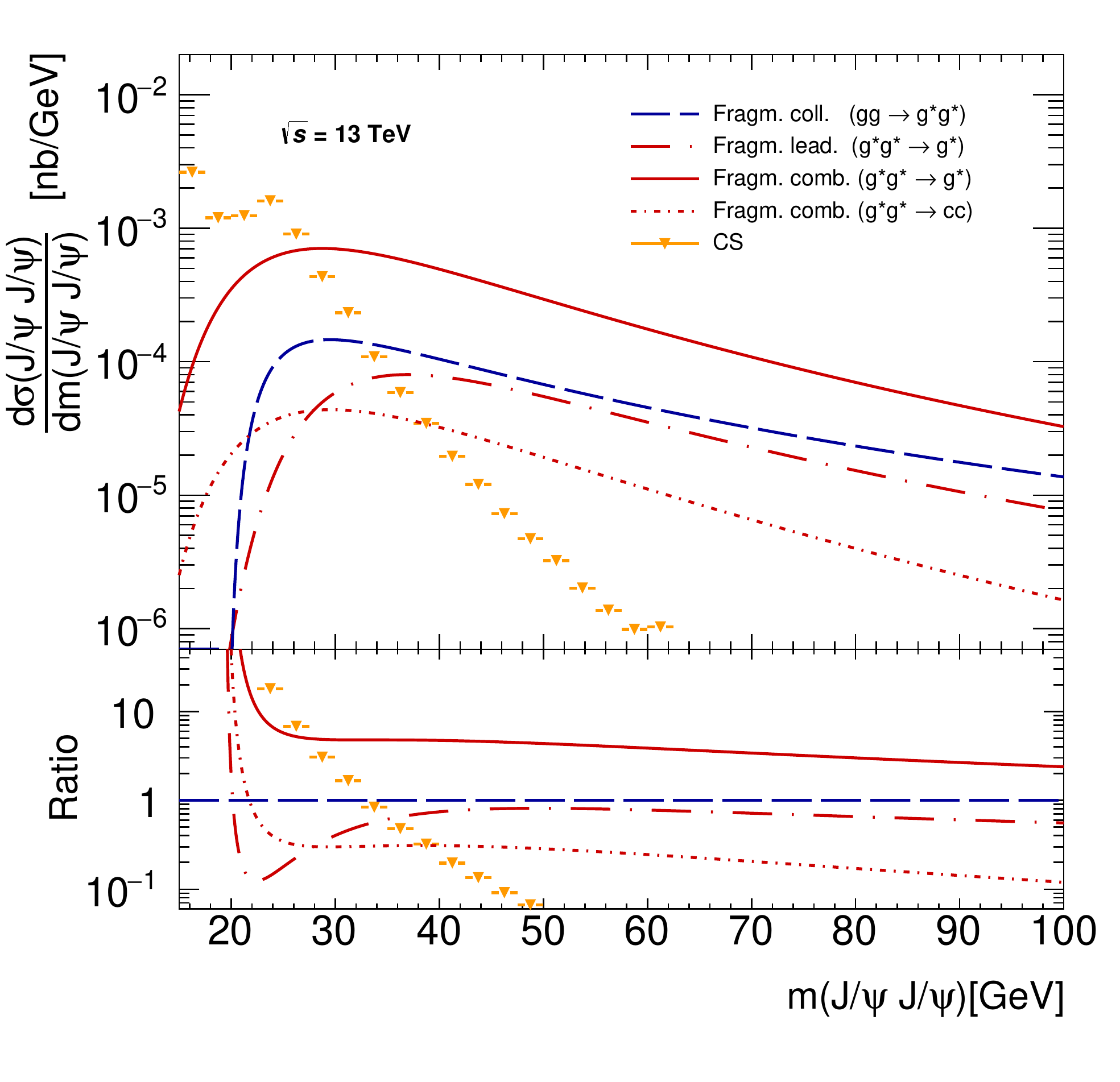}
\includegraphics[width=6.0cm]{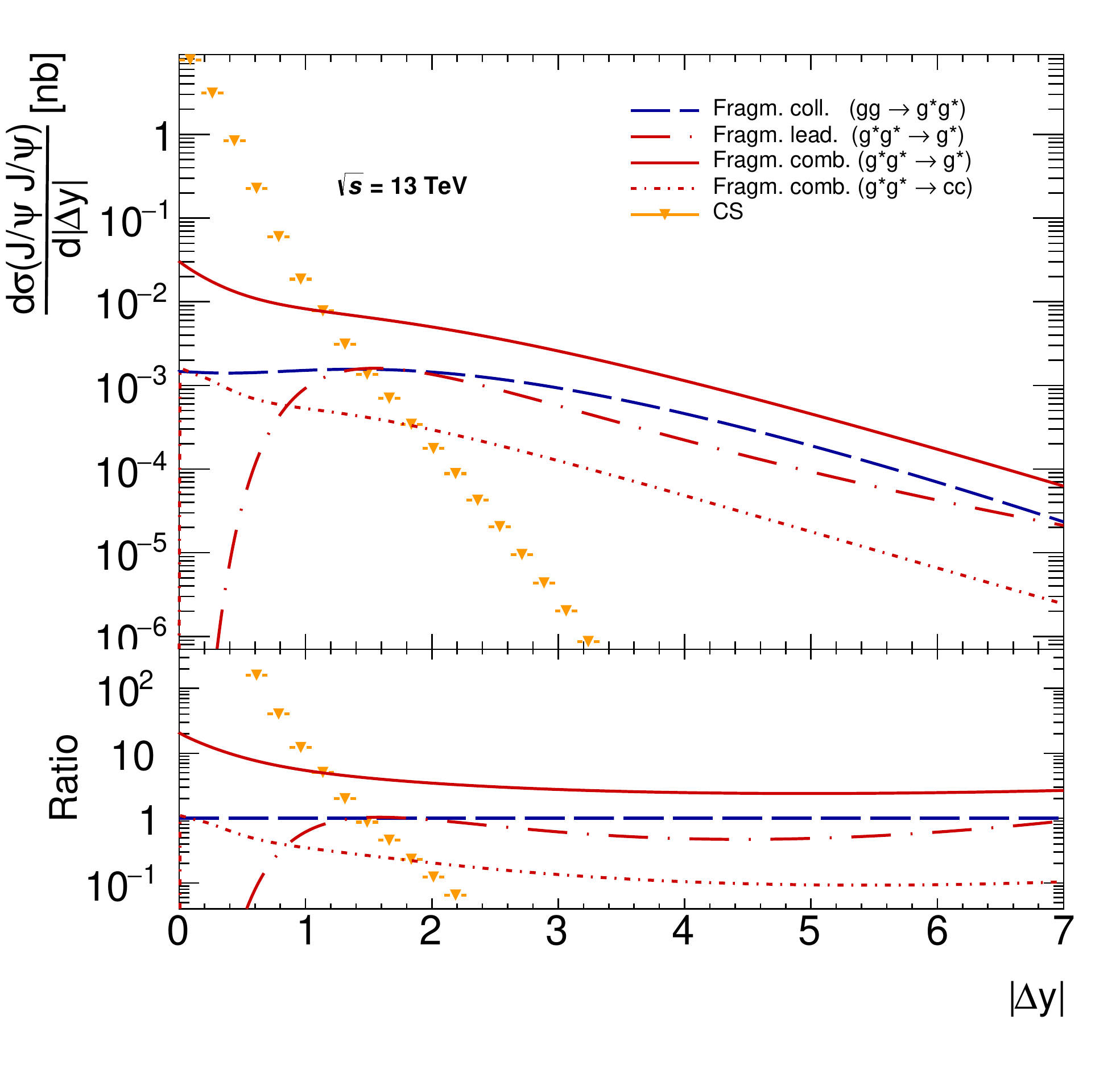}
\caption{Different contributions to the double $J/\psi$ production
calculated as functions of invariant mass $m(J/\psi, J/\psi)$, 
transverse momentum of leading $J/\psi$ meson $p_T^{\rm lead}(J/\psi)$ and rapidity
separation $|\Delta y(J/\psi, J/\psi)|$ at $\sqrt s = 13$~TeV.
The kinematical cut $p_T(J/\psi) > 10$~GeV is applied for both $J/\psi$ mesons.
The A0 gluon distribution in proton is used.}
\label{fig_contrib}
\end{center}
\end{figure}

\begin{figure}
\begin{center}
\includegraphics[width=6.0cm]{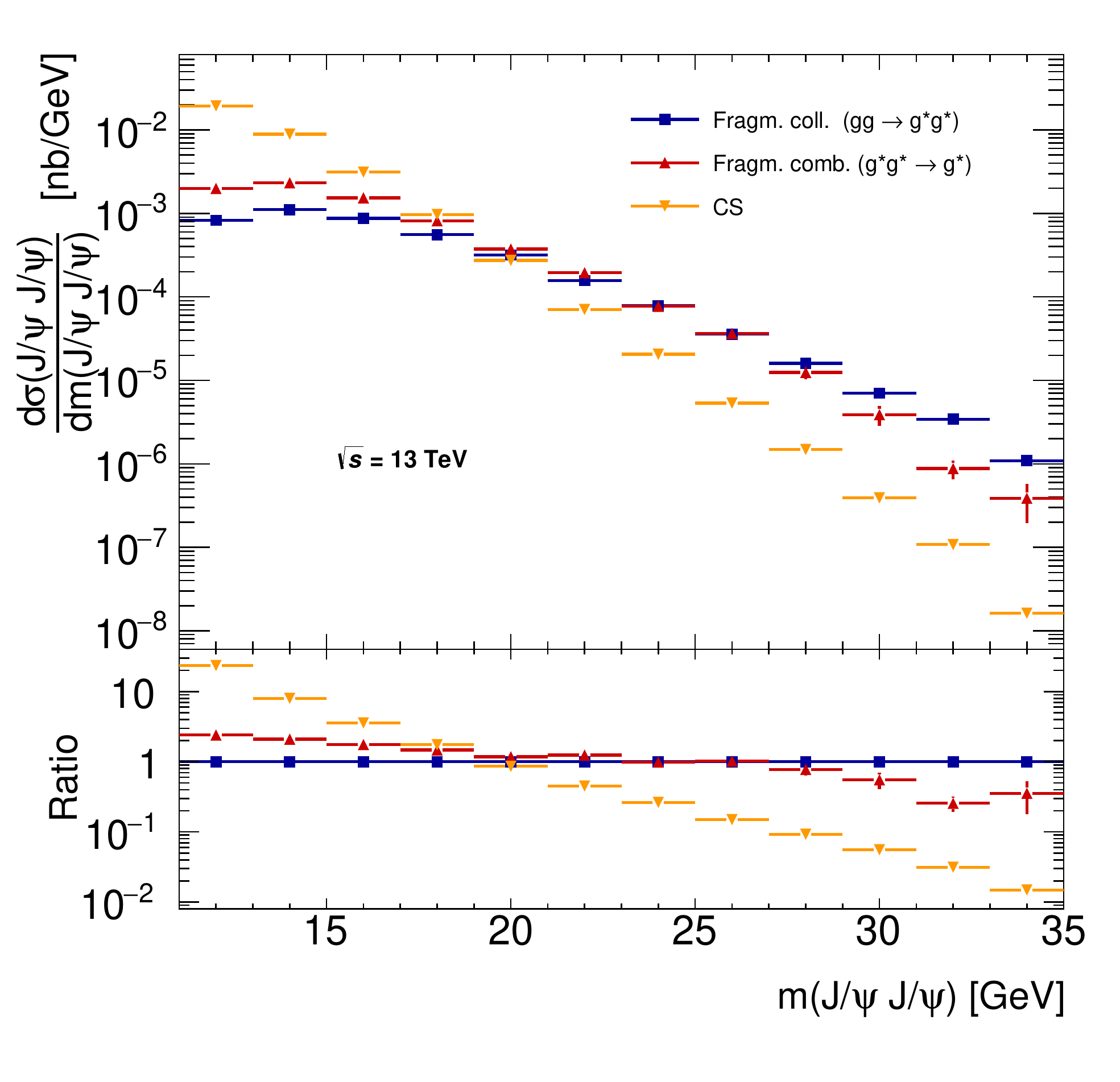}
\includegraphics[width=6.0cm]{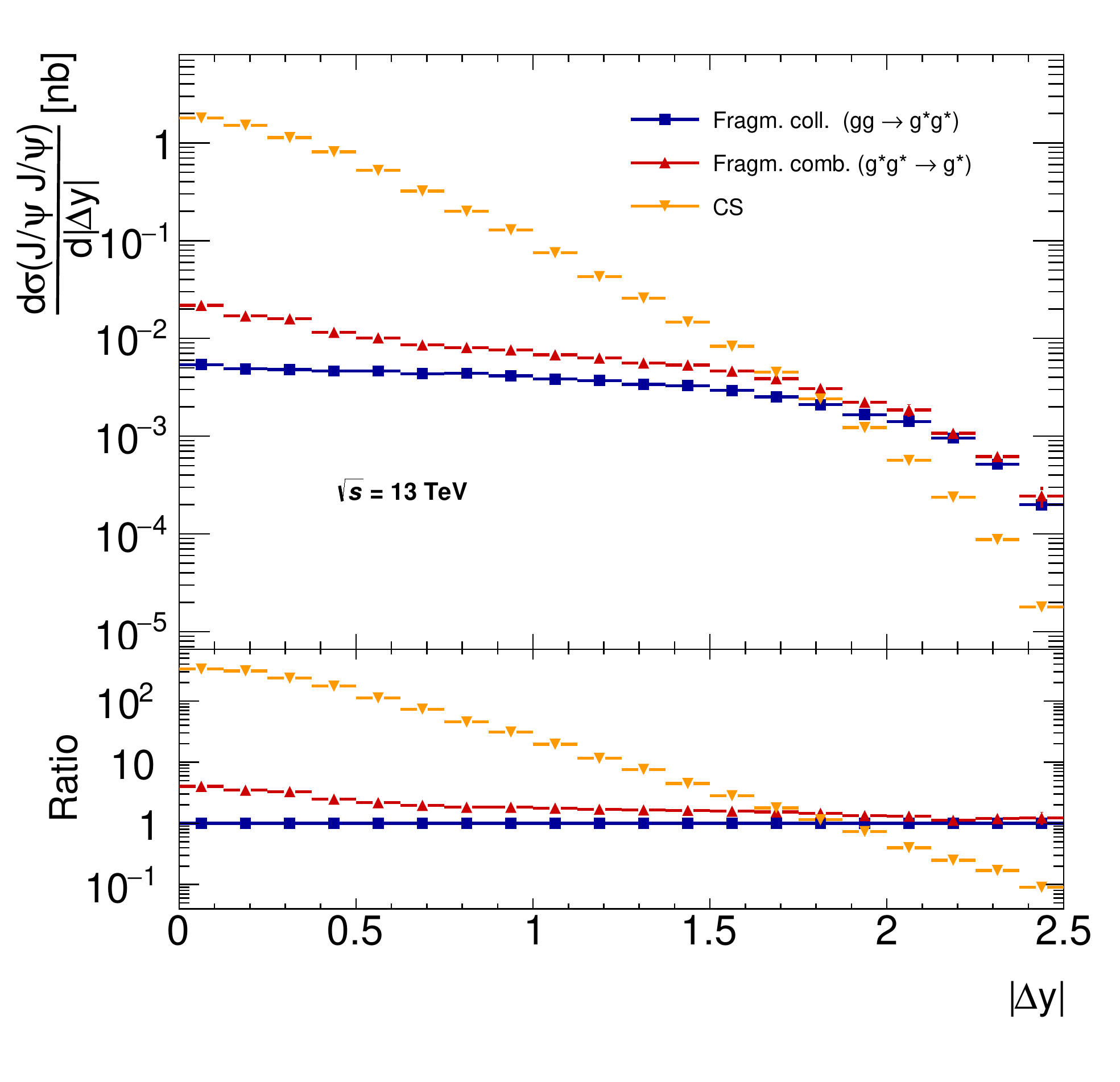}
\caption{Different contributions to the double $J/\psi$ production
calculated as functions of invariant mass $m(J/\psi, J/\psi)$ and rapidity
separation $|\Delta y(J/\psi, J/\psi)|$ at $\sqrt s = 13$~TeV.
The kinematical cuts $4.5 < p_T(J/\psi) < 10$~GeV and $2 < y(J/\psi) < 4$ 
are applied for both $J/\psi$ mesons.
The A0 gluon distribution in proton is used.}
\label{fig_contribLHCb}
\end{center}
\end{figure}

In Fig.~\ref{fig_contrib} we show the differential cross sections of double ${J/\psi}$ 
production calculated as a functions of $J/\psi$ pair invariant mass $m(J/\psi, J/\psi)$ and 
difference in rapidity between the $J/\psi$ mesons $|\Delta y(J/\psi, J/\psi)|$ 
at $\sqrt s = 13$~TeV.
We have required $p_T(J/\psi) > 10$~GeV for both produced mesons,
that ensures the validity of the fragmentation approach used.
Moreover, this restriction close to the CMS or ATLAS conditions.
One can see that an accurate account of combinatorial
contributions, originated from the cascade gluon fragmentation into
the $J/\psi$ mesons (labeled as "fragm. comb."), significantly (up to an order of magnitude) 
increase the cross section compared to the 
single gluon fragmentation, governed by the LO gluon-gluon
fusion subprocess\footnote{Here we have applied the MMHT'2014 (LO) parton density set\cite{62}.}
(labeled as "fragm. coll.").
For the latter, we reproduce the results\cite{27}.
To highlight the importance of the combinatorial gluon fragmentation, 
we show the results obtained using the simplified selection of ${J/\psi}$ pair in each event,
where one of the $J/\psi$ mesons is originated from the gluon produced in the 
hard scattering subprocess and another one is produced from the leading cascade gluon (labeled as "fragm. lead.").
This selection criterion almost corresponds to the 
collinear limit, as it is clearly demonstrated in Fig.~\ref{fig_contrib}.
Next, we find that the cascade gluon fragmentation plays a dominant 
role at large invariant masses $m(J/\psi, J/\psi) \geq 25$~GeV
and $|\Delta y(J/\psi, J/\psi)| \geq 1$,
where it greatly overestimates the CS contributions.
Taking into account these combinatorial contributions
results in the 
drastical rise of the double $J/\psi$ production cross sections 
at large $m(J/\psi, J/\psi)$, 
where the strong discrepancy between 
the NRQCD estimations (including both the CS and CO terms) 
and experimental data, taken by the CMS and ATLAS Collaborations, is observed.
Contrary, the combinatorial fragmentation effects should be significantly less pronounced
at forward rapidities, which are covered by the LHCb measurements.
To demonstrate it, we have repeated the calculations 
under the requirements $4.5 < p_T(J/\psi) < 10$~GeV and $2 < y(J/\psi) < 4.5$.
The upper limit of $p_{T}(J/\psi)$ is set to be the same as in LHCb analyses\cite{3,4} 
while lower limit corresponds to the region, where the fragmentation approach is 
valid.
Our results for distributions in 
$m(J/\psi,J/\psi)$ and $|\Delta y(J/\psi,J/\psi)|$
are shown in Fig.~\ref{fig_contribLHCb}.
One can see that the cascade gluon fragmentation 
gives only small contribution to the forward $J/\psi$ pair 
production and, in principle, can be safely neglected. 
It can be easily understood 
since at large rapidities (or, equivalently, at
large momentum fraction $x$ of one of the interacting gluons) the gluon 
emissions in the initial state are insufficient. 

Concerning the contributions from charm fragmentation,
their role (compared to the LO predictions of conventional pQCD) is also a
bit enhanced due to the multiple gluon emissions in
the initial state. 
We find that these processes amount several percent of 
the ${J/\psi}$ pair production cross section 
(see Figs.~\ref{fig_contrib} and \ref{fig_contribLHCb}) and, of course, can be considered as additional non-leading
terms\footnote{To generate $c\bar c$ events 
in the off-shell gluon-gluon fusion the Monte-Carlo event generator \textsc{pegasus}\cite{55} has been used.}.

Thus, we have shown that taking into account the combinatorial contributions
from the cascade gluon fragmentation
could fill the gap between the NRQCD
predictions and experimental data. 
However, to perform the quantitative comparison with the available CMS\cite{1} and ATLAS\cite{2} measurements
one has to include into the analysis a number 
of other possible fragmentation channels
playing role at low and moderate transverse momenta.
Moreover, additional feeddown contributions 
to the double $J/\psi$ production 
from the $\chi_c$ and $\psi^\prime$ decays should be taken into account.
An accurate theoretical description
requires a rather long-time numerical calculations. 
So, here we only claim the possible importance of the combinatorial
fragmentation terms above and left their 
further cumbersome analysis for a forthcoming dedicated study.

Now we turn to available LHCb data
collected at $\sqrt s = 7$ and $13$~TeV\cite{3,4}. 
These data refer to $p_T(J/\psi) < 10$~GeV, $m(J/\psi, J/\psi) < 15$~GeV 
and forward rapidity region, $2 < y(J/\psi) < 4.5$.
Since the combinatorial contributions from gluon and/or charmed 
quark fragmentation are almost negligible there,
only the CS terms and DPS production mechanism play the role.
The latter give us the possibility to easily 
extract the key parameter of DPS mechanism, the effective cross section $\sigma_{\rm eff}$, 
from the LHCb measurements.
The feeddown contributions from radiative $\chi_{c}$ and $\psi^\prime$ decays 
to the SPS cross section, which is governed by the subprocess~(1),
are also unimportant 
at small transverse momenta 
and invariant mass $m(J/\psi, J/\psi)$, see discussions\cite{37,63}.
Thus, we neglect below all these terms for simplicity.
To evaluate the DPS contribution
to the $J/\psi$ pair production
we use the results of our previous studies
and strictly follow the approach\cite{19}
for the inclusive cross section
$\sigma(pp \to J/\psi + X)$, entering to the DPS factorization formula~(9).
So, the determination of $\sigma_{\rm eff}$ can be performed 
in a self-consistent way.

The following kinematical variables have been investigated in the LHCb analyses\cite{3,4}:
transverse momentum $p_T(J/\psi, J/\psi)$, rapidity $y(J/\psi,J/\psi)$ and invariant mass of the $J/\psi$ pair,
transverse momentum and rapidity of $J/\psi$ mesons, 
differences in the azimuthal angle $|\Delta\phi(J/\psi, J/\psi)|$
and rapidity $|\Delta y(J/\psi, J/\psi)|$
between the produced mesons and 
transverse momentum asymmetry ${\cal A}_T$, defined as
\begin{gather}
  {\cal A}_T = \left| {p_T(J/\psi_1) - p_T(J/\psi_2) \over p_T(J/\psi_1) + p_T(J/\psi_2)} \right|.
\end{gather}
\noindent
The measurements have been performed for $p_T(J/\psi,J/\psi) > 1$~GeV, $p_T(J/\psi,J/\psi) > 3$~GeV
and in the whole $p_T(J/\psi,J/\psi)$ range.
We consider $\sigma_{\rm eff}$ as an independent parameter
and perform a simultaneous fit to the LHCb data.
The fitting procedure was separately done for each of the measured kinematical 
distributions employing the fitting algorithm as implemented in the commonly used \textsc{gnuplot} package\cite{64}.

Not all of the existing data sets are equally informative for the $\sigma_{\rm eff}$ extraction. 
Using the data where the DPS contribution is smaller 
than the uncertainty of the "main" contribution would only increase the total 
error. So, our fit is based on the following distributions (all measured at
$\sqrt s = 7$~TeV and $13$~TeV):
single $J/\psi$ transverse momentum $p_T(J/\psi)$;
single $J/\psi$ rapidity $y(J/\psi)$;
invariant mass $m(J/\psi,J/\psi)$;
transverse momentum of $J/\psi$ pair;
rapidity of $J/\psi$ pair;
transverse momentum asymmetry ${\cal{A}}_T$;
rapidity separation between the two $J/\psi$ mesons $|\Delta y(J/\psi,J/\psi)|$.
For all observables except $|\Delta y(J/\psi,J/\psi)|$ we used the data 
without cuts on $p_T(J/\psi,J/\psi)$ and with $p_T(J/\psi,J/\psi) > 1$~GeV,
while for the rapidity separation $|\Delta y(J/\psi,J/\psi)|$ we used the sets
without cuts on $p_T(J/\psi,J/\psi)$, with $p_T(J/\psi,J/\psi) > 1$~GeV, and with 
$p_T(J/\psi,J/\psi) > 3$~GeV.

The obtained mean-square average of the fitted values is $\sigma_{\rm eff} = 17.5 \pm 4.1$~mb,
where corresponding uncertainty is estimated in the conventional way using Student's 
t-distribution at the confidence level $P = 95$\%.
Here we achieve a remarkable agreement 
with the majority of other $\sigma_{\rm eff}$ estimations  
based on different final states, such as, for example, $W + 2$~jets\cite{65,66},
$2\,\gamma + 2$~jets\cite{67}, $\gamma + 3$~jets\cite{68}, 
$4$~jets\cite{38}, $J/\psi + D^+$, $J/\psi + D^0$, $J/\psi + \Lambda_c^+$\cite{69}, 
$\Upsilon(1S) + D^0$\cite{39}. Thus, our result  
supports the expectation about the universality of this parameter for a wide
range of processes with essentially different kinematics, energies and hard scales.
The obtained value of $\sigma_{\rm eff}$ significantly exceeds previous
estimations based on the same final state, $J/\psi + J/\psi$,
which are typically of about $2 - 5$~mb\cite{42,43,44}.
Of course, the results\cite{42,43,44} also contradict to the most of the measured $\sigma_{\rm eff}$
values\cite{65,66,67,68,69}.

\begin{figure}
\begin{center}
\includegraphics[width=6.0cm]{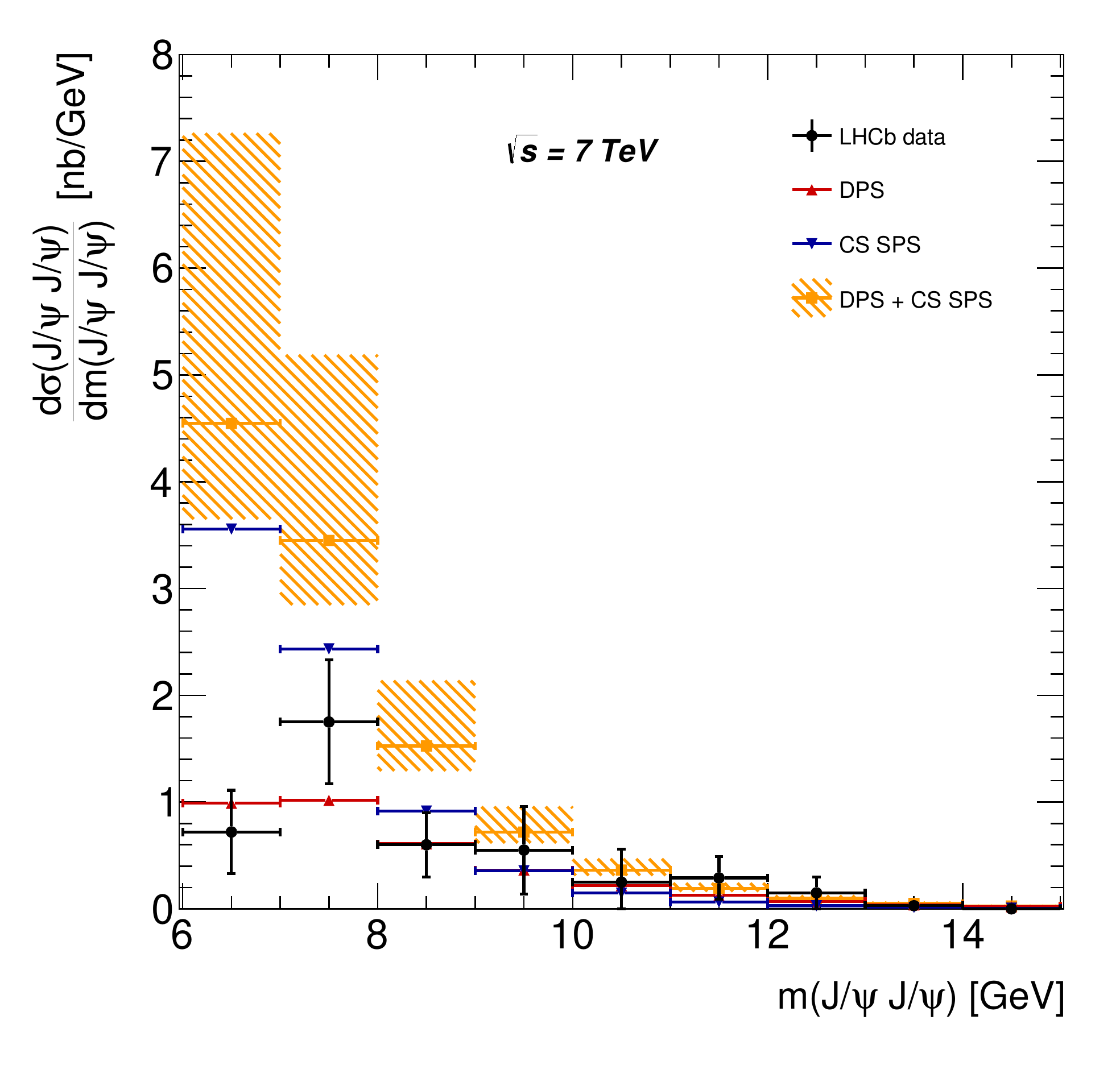}
\includegraphics[width=6.0cm]{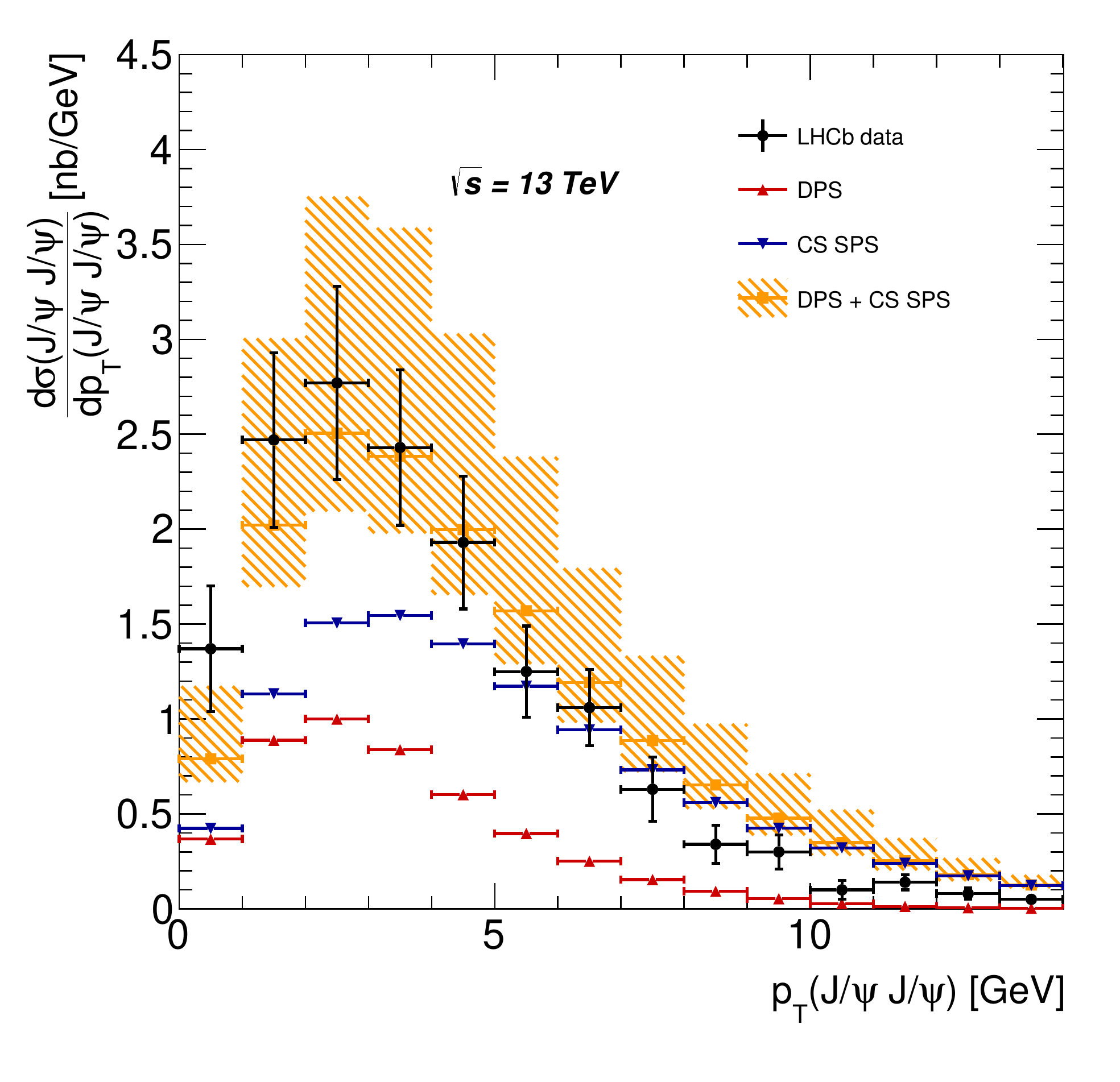}
\caption{Differential cross sections of double $J/\psi$ production 
as functions of invariant mass $m(J/\psi, J/\psi)$ and 
transverse momentum $p_T(J/\psi,J/\psi)$ 
calculated at $\sqrt s = 7$~TeV (left panel) and $\sqrt s = 13$~TeV (right panel).
The kinematical cuts applied are described in the text.
The A0 gluon distribution in proton is used.}
\label{fig_LHCba}
\end{center}
\end{figure}

\begin{figure}
\begin{center}
\includegraphics[width=6.0cm]{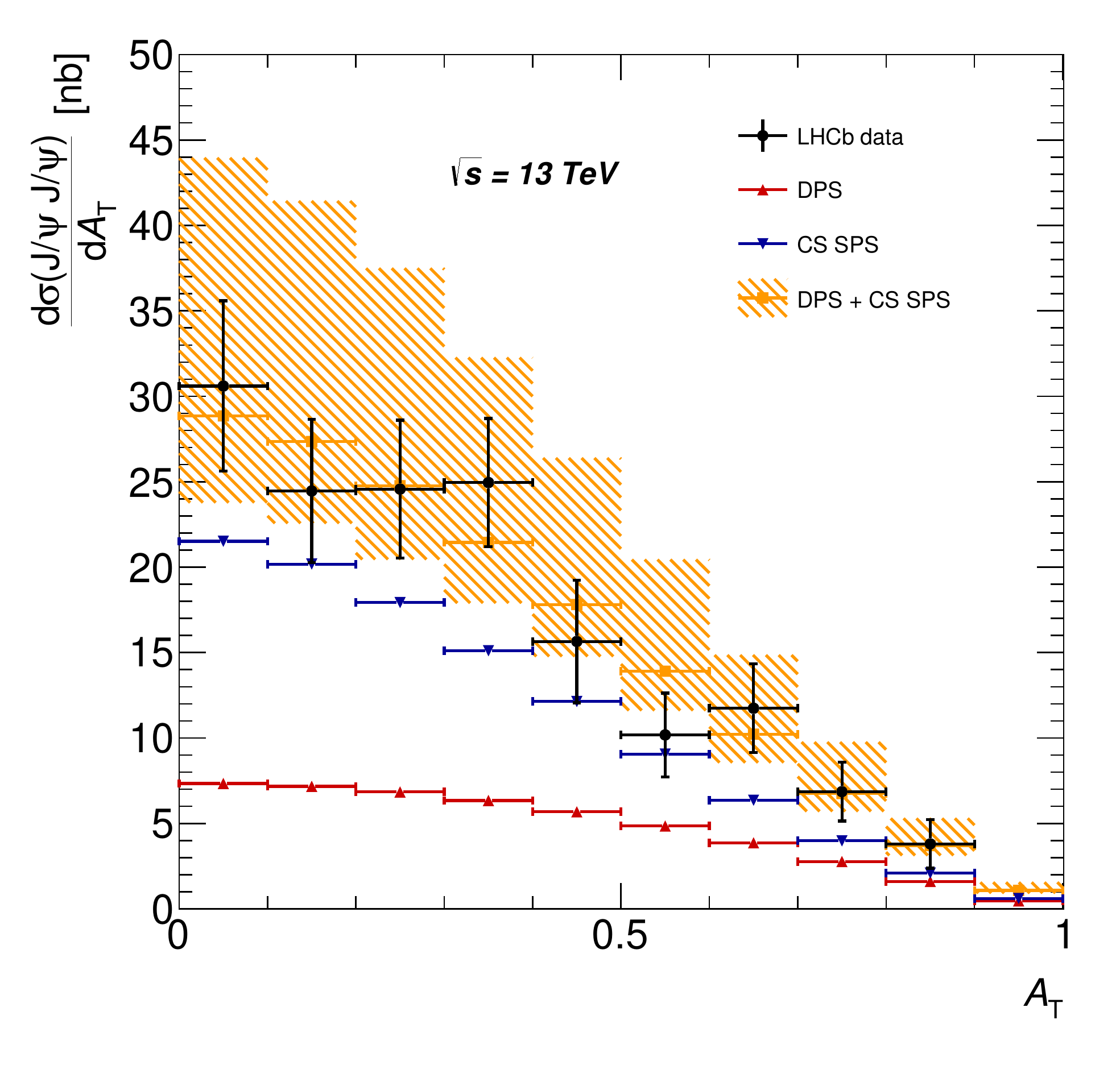}
\includegraphics[width=6.0cm]{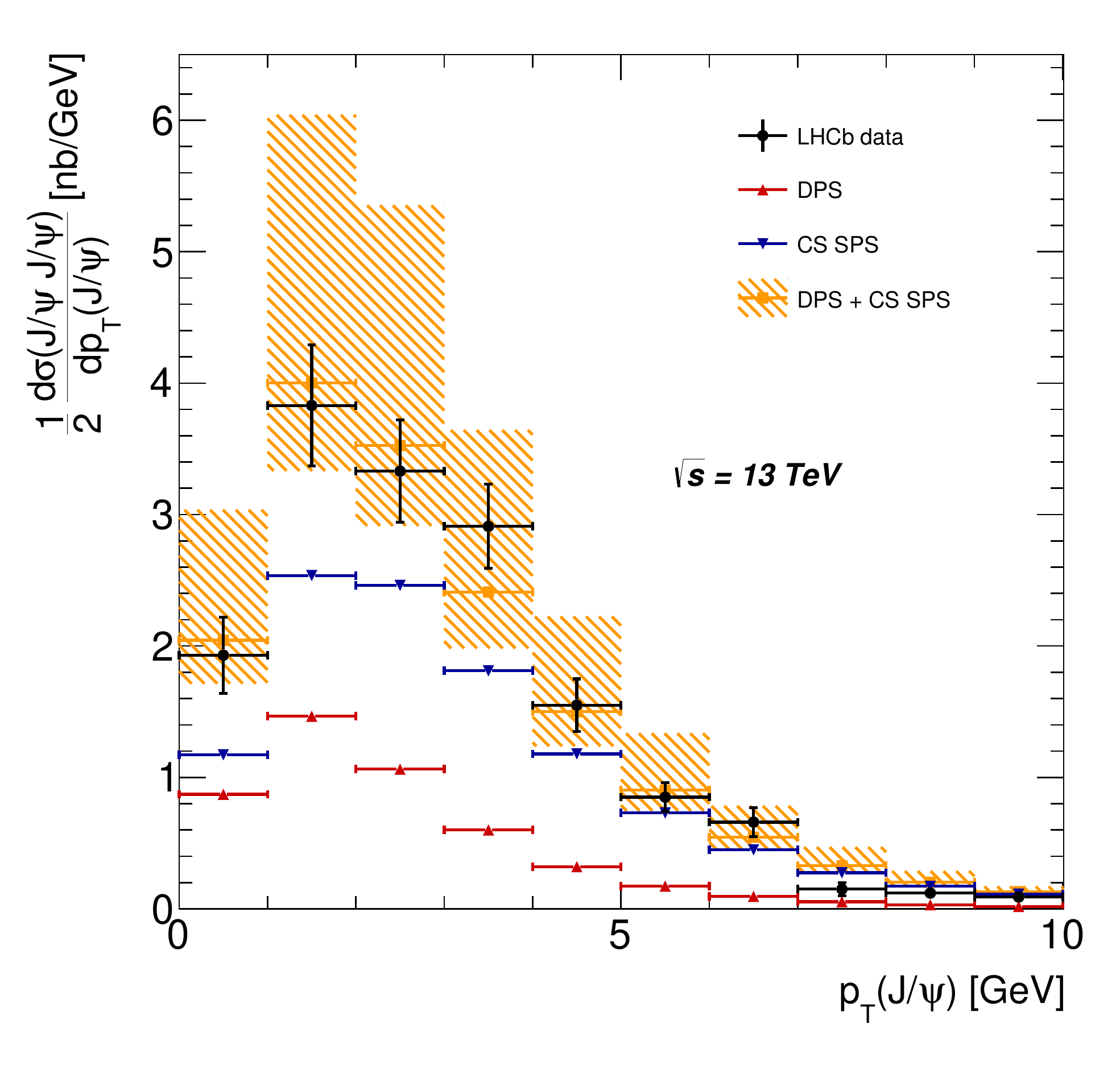}
\includegraphics[width=6.0cm]{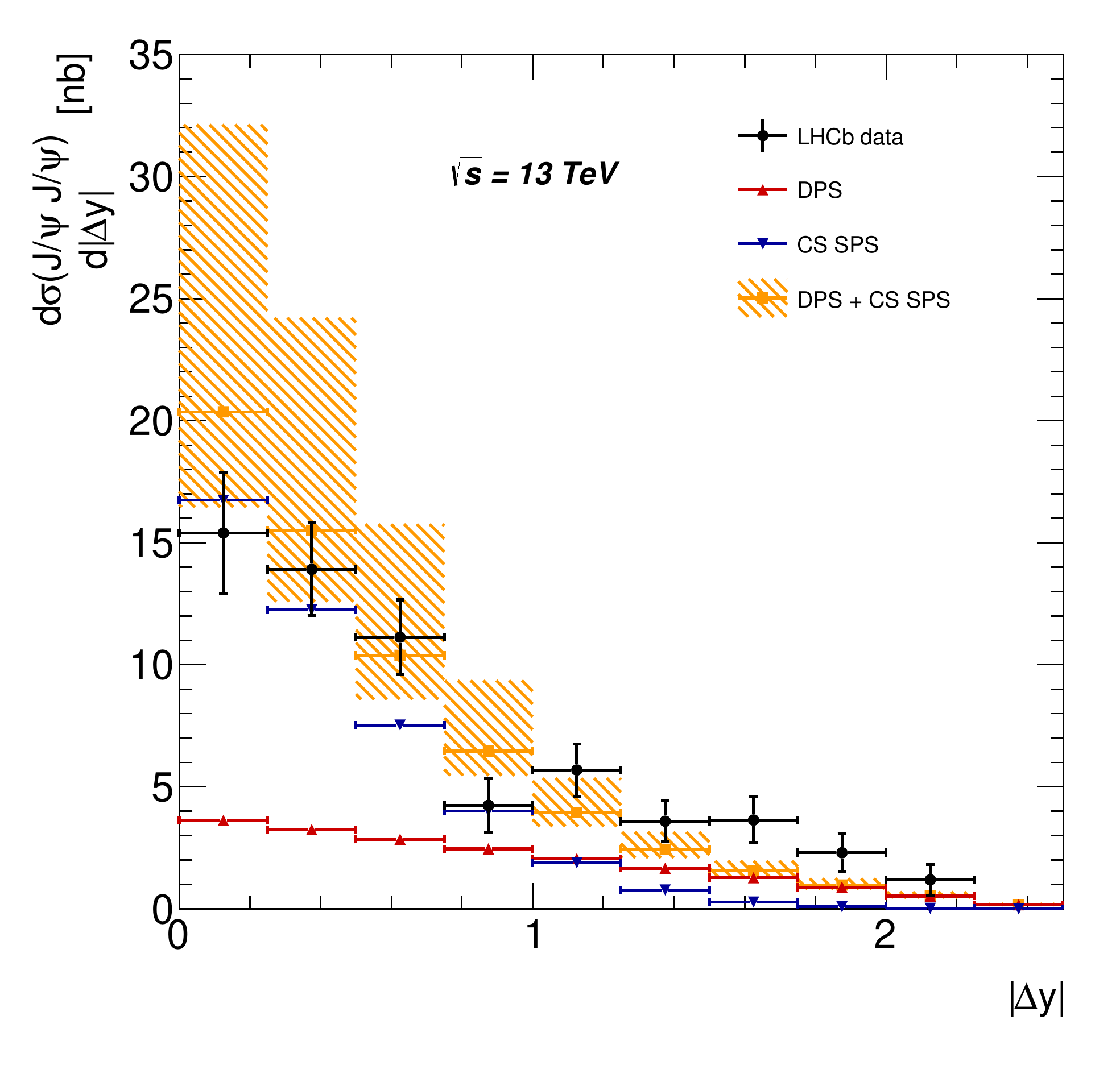}
\includegraphics[width=6.0cm]{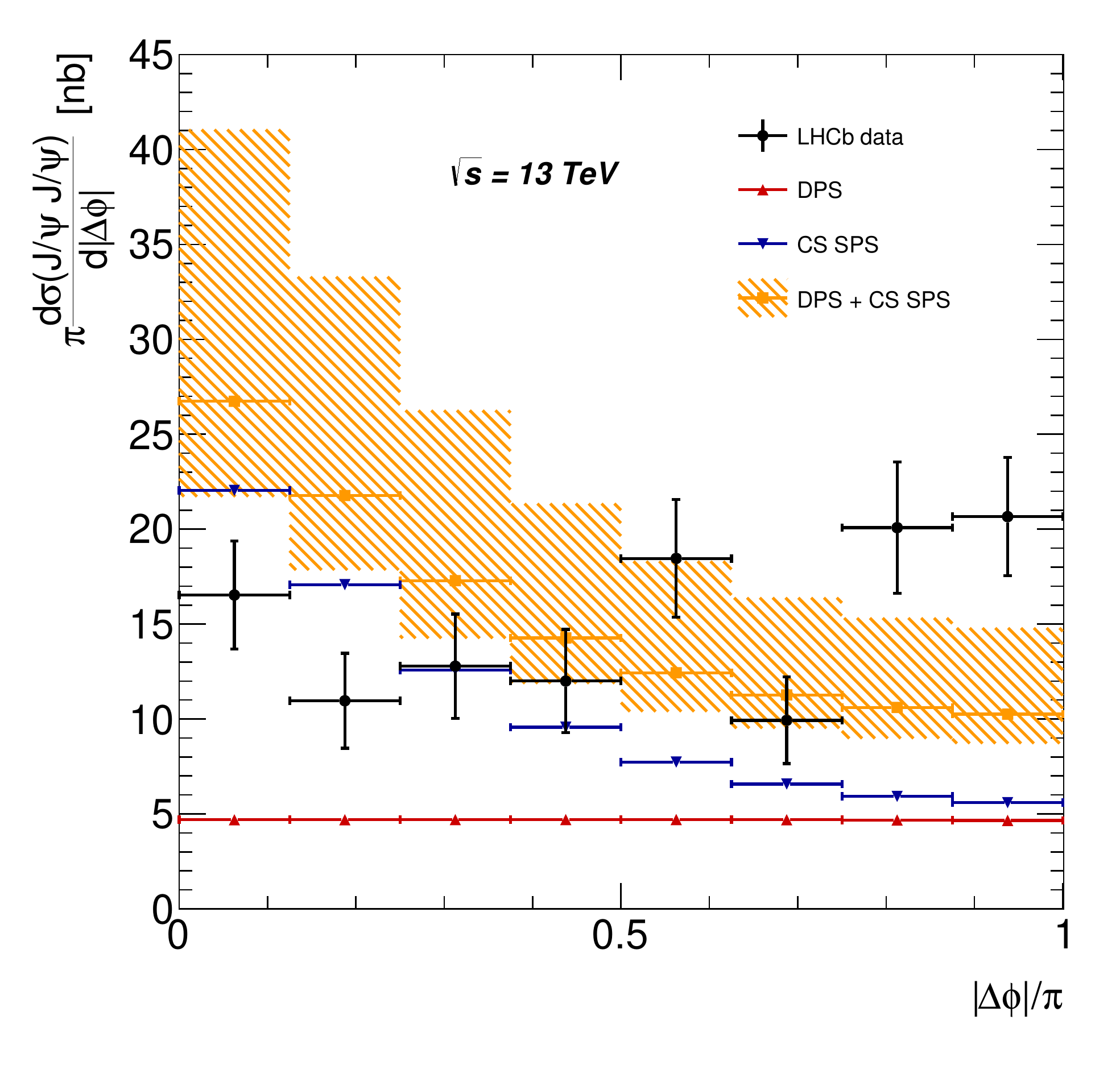}
\includegraphics[width=6.0cm]{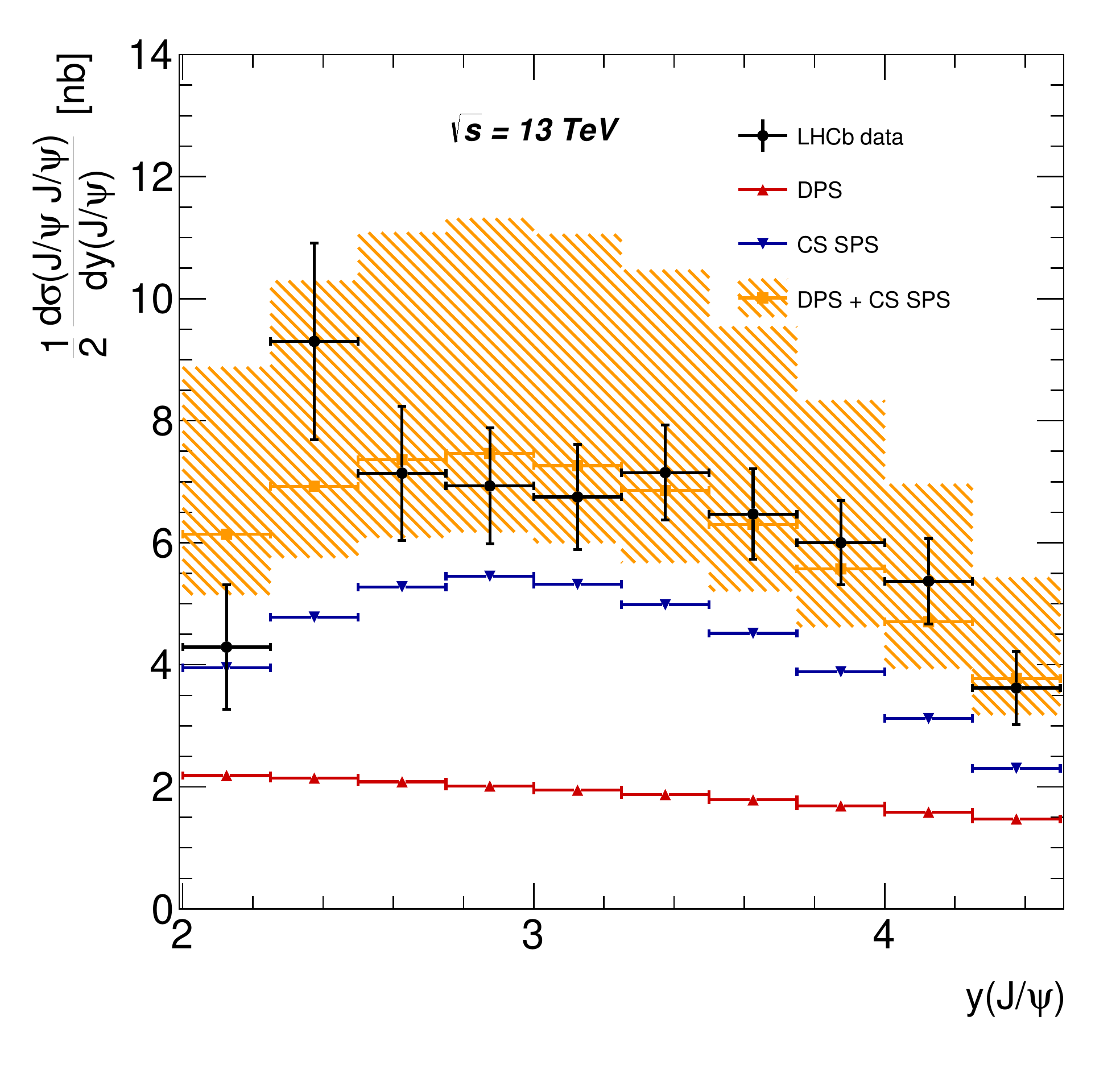}
\includegraphics[width=6.0cm]{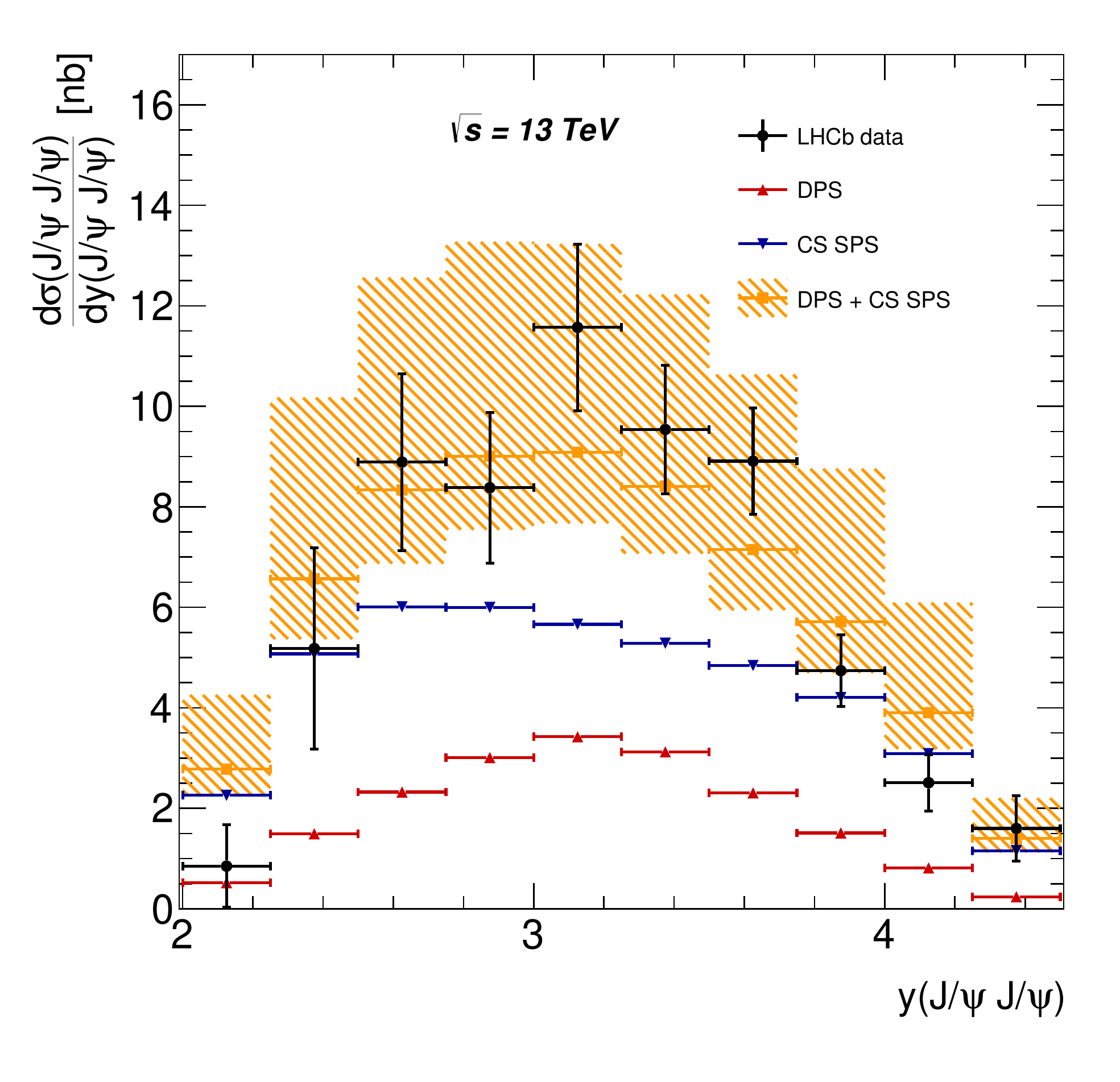}
\includegraphics[width=6.0cm]{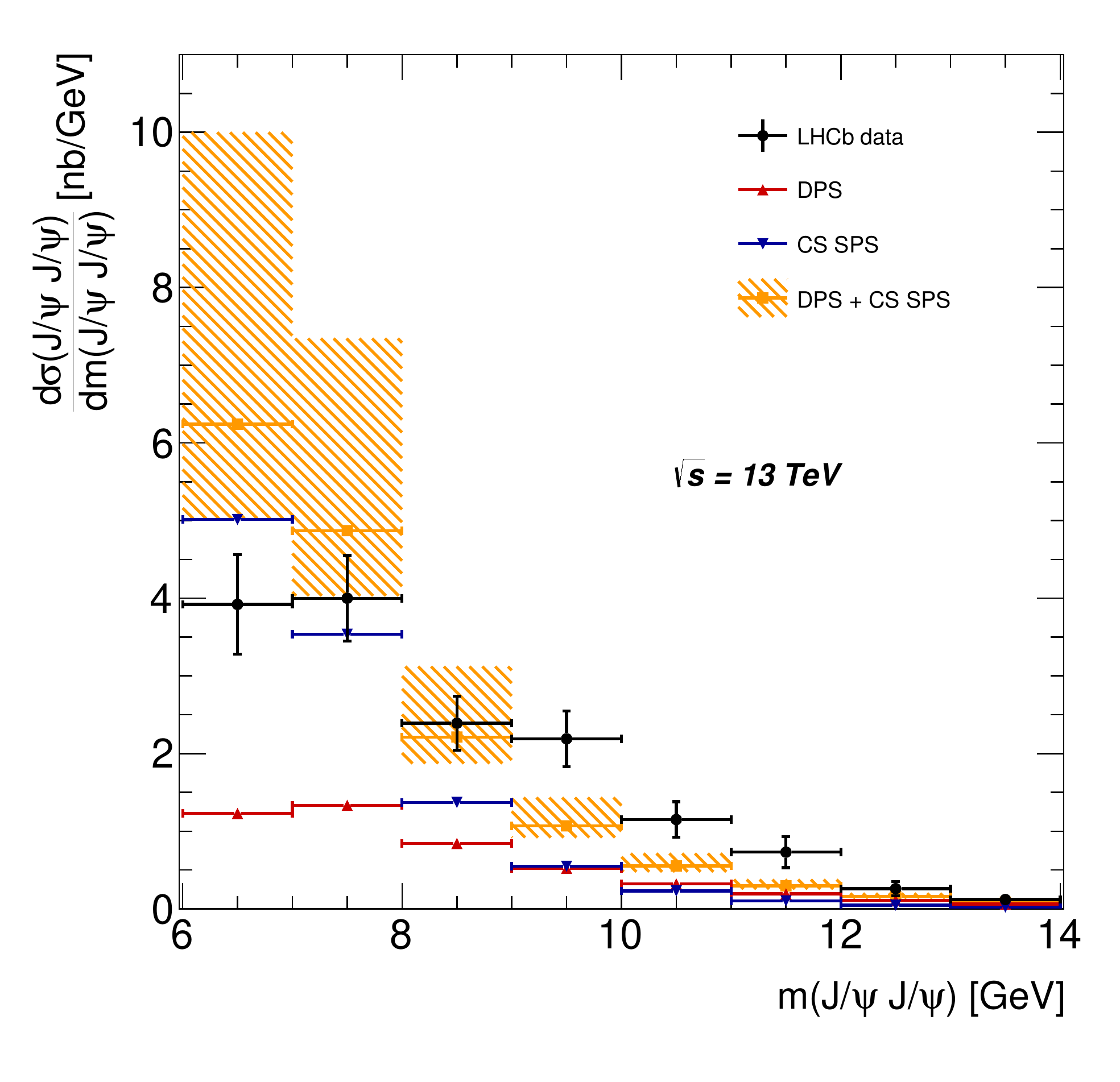}
\caption{Prompt double $J/\psi$ production 
as functions of different kinematical variables 
calculated at $\sqrt s = 13$~TeV.
The kinematical cuts applied are described in the text.
The A0 gluon distribution in proton is used.}
\label{fig_LHCb1}
\end{center}
\end{figure}

\begin{figure}
\begin{center}
\includegraphics[width=6.0cm]{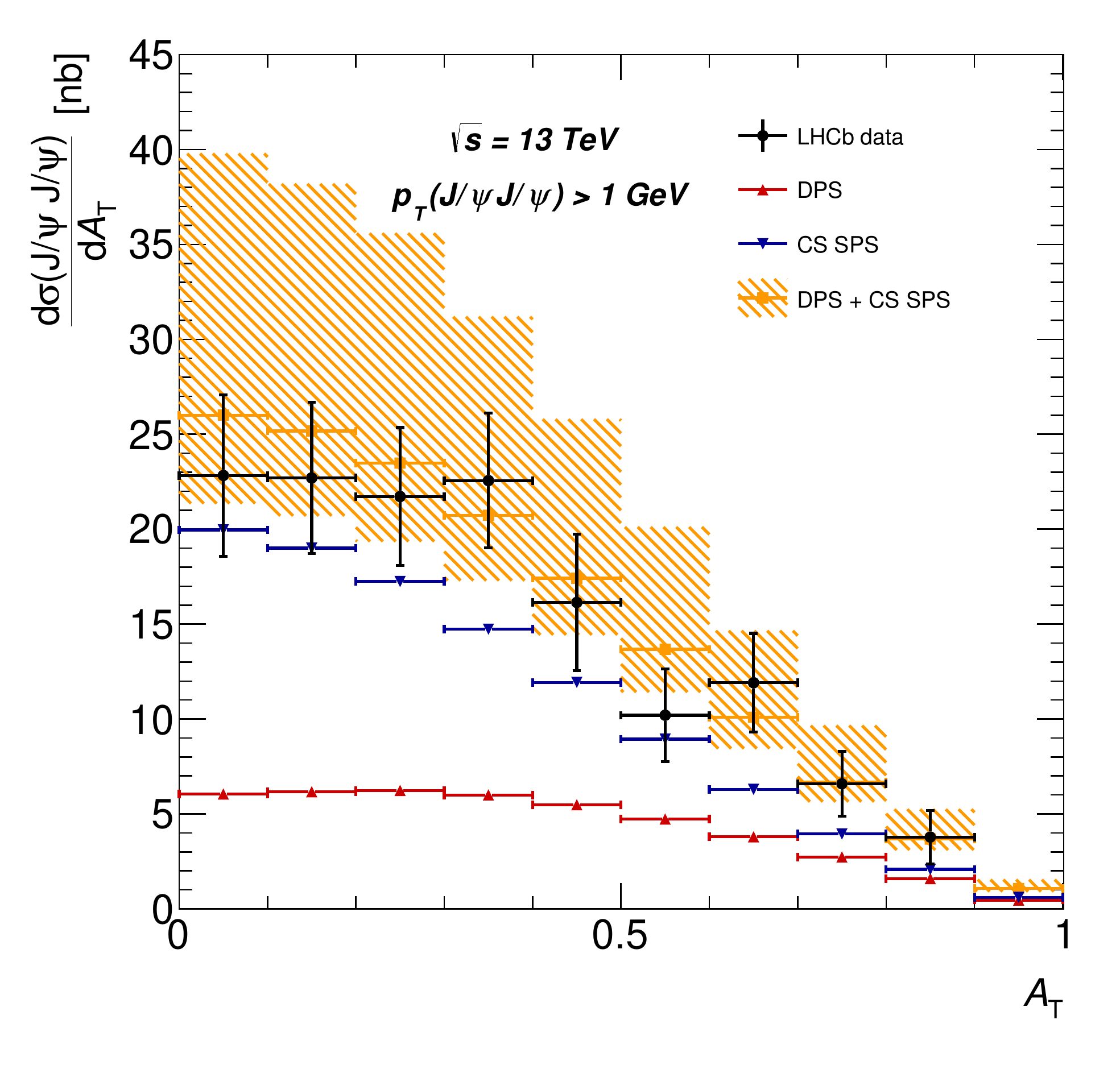}
\includegraphics[width=6.0cm]{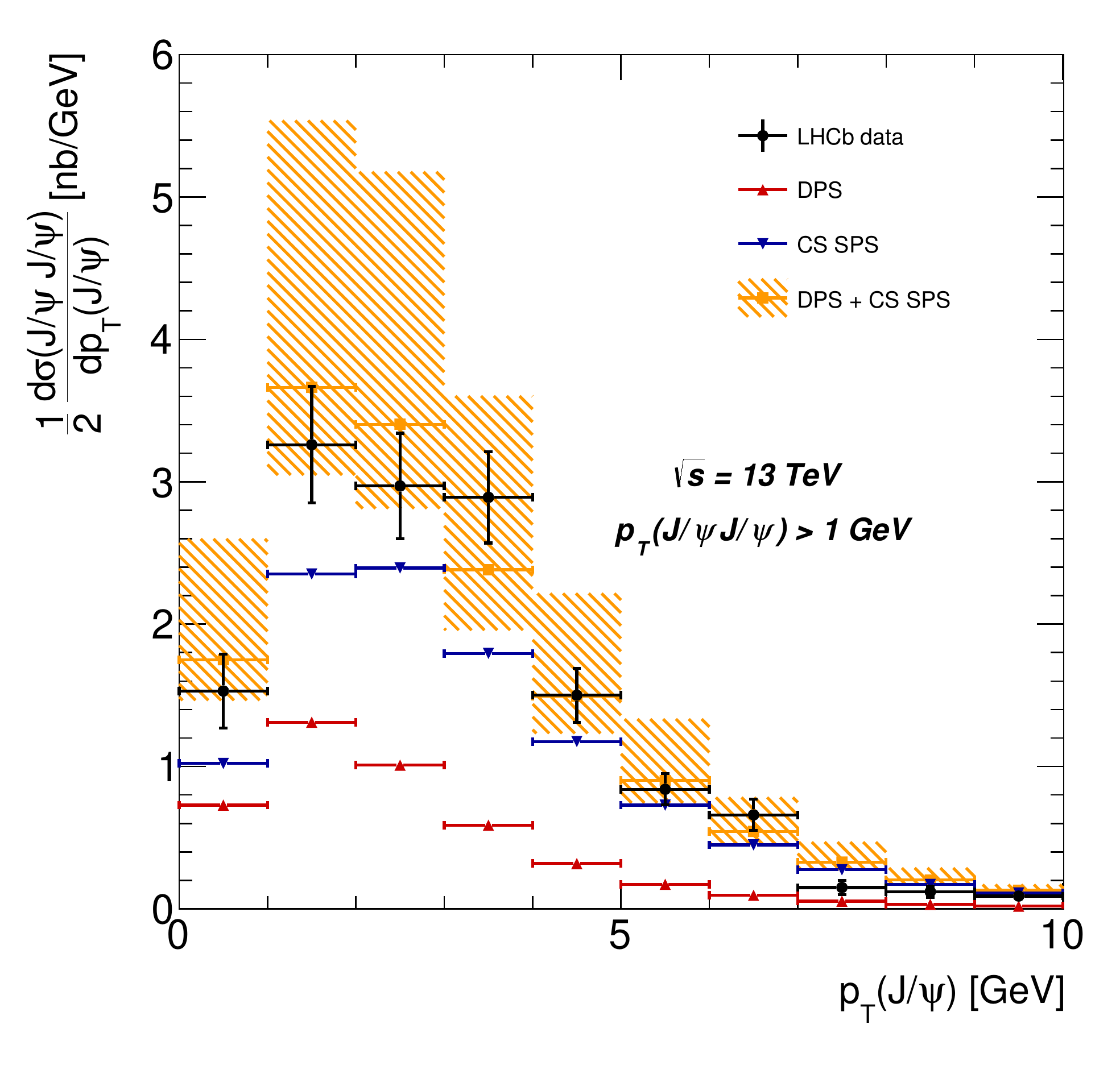}
\includegraphics[width=6.0cm]{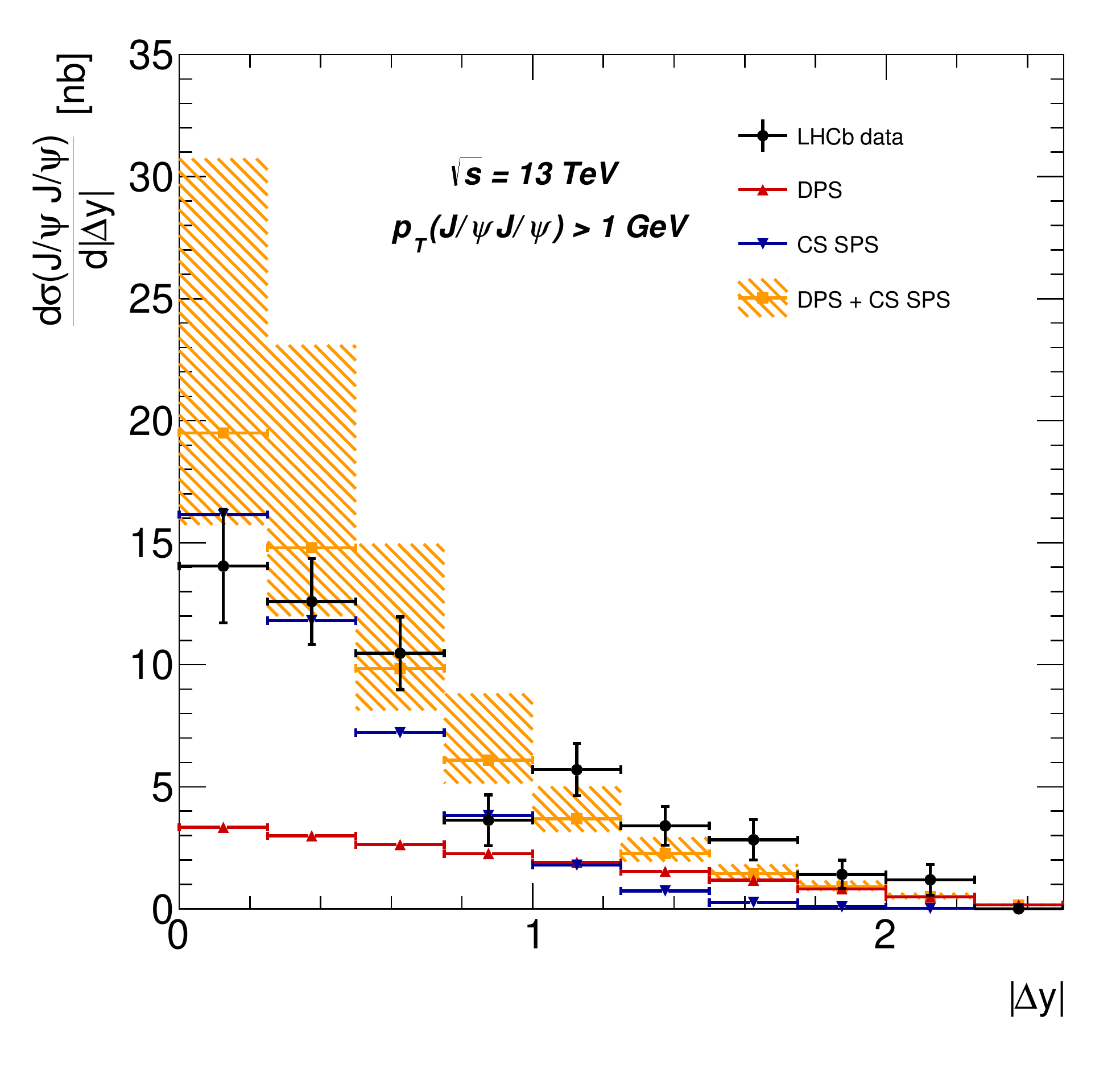}
\includegraphics[width=6.0cm]{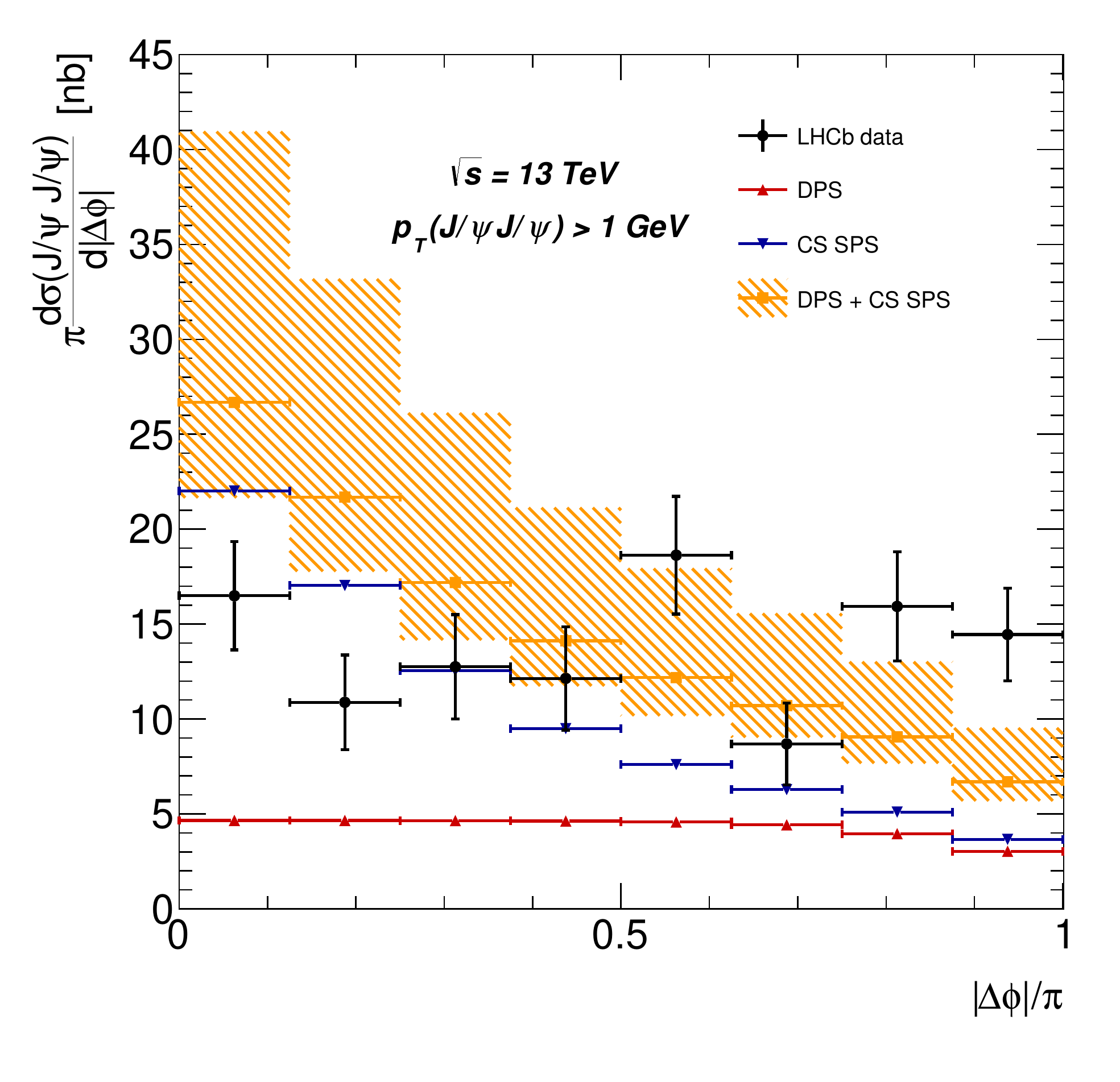}
\includegraphics[width=6.0cm]{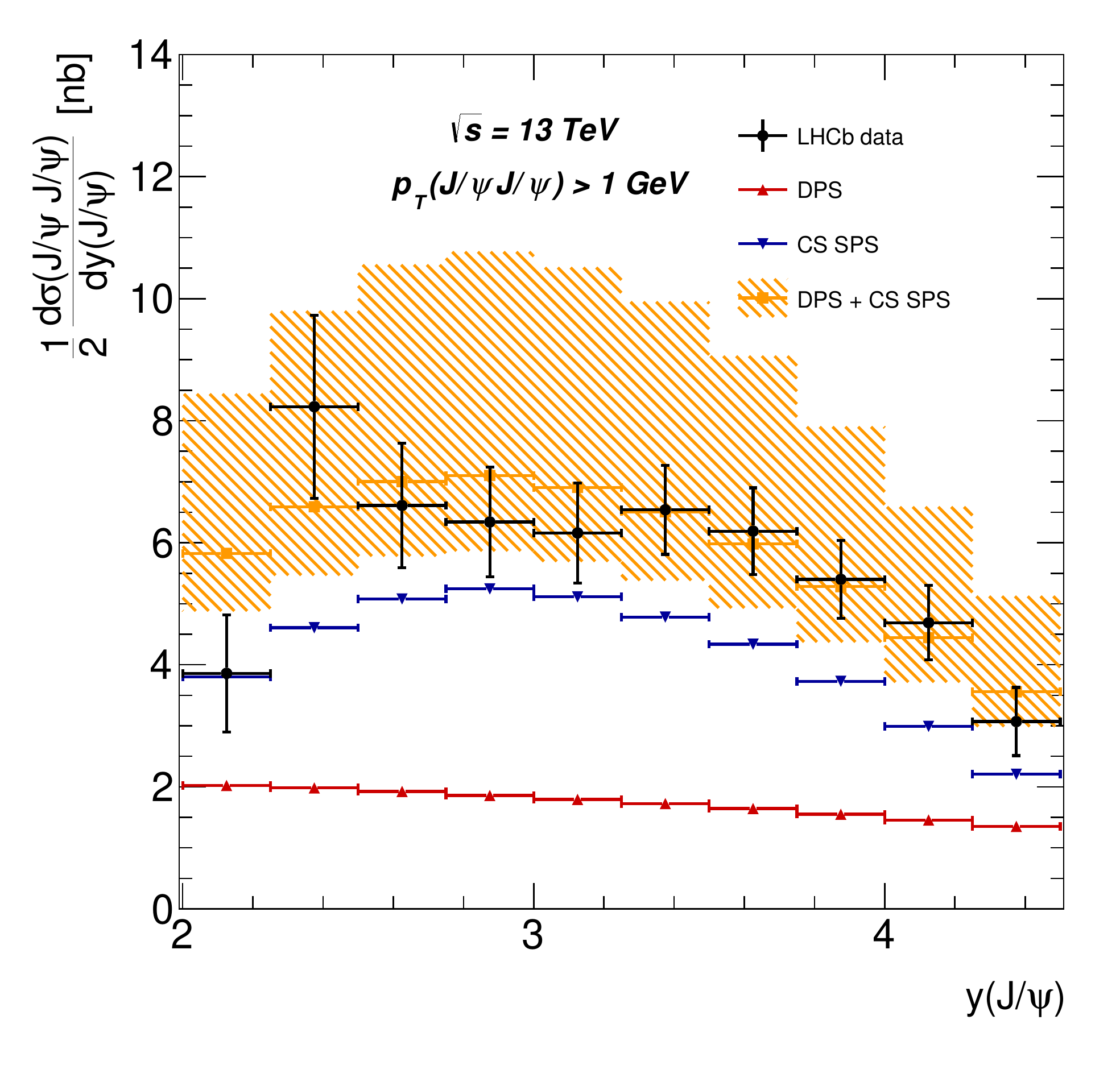}
\includegraphics[width=6.0cm]{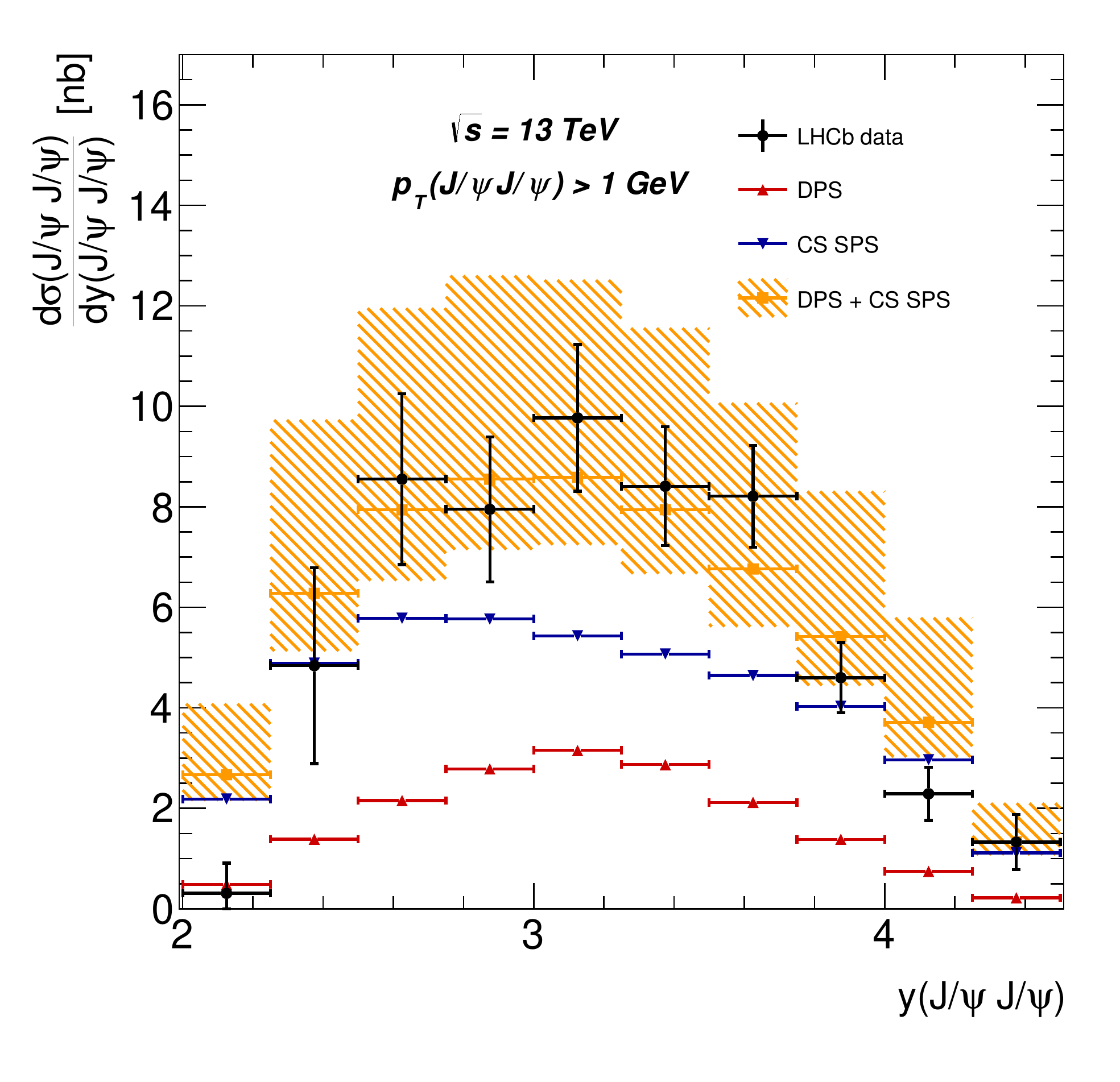}
\includegraphics[width=6.0cm]{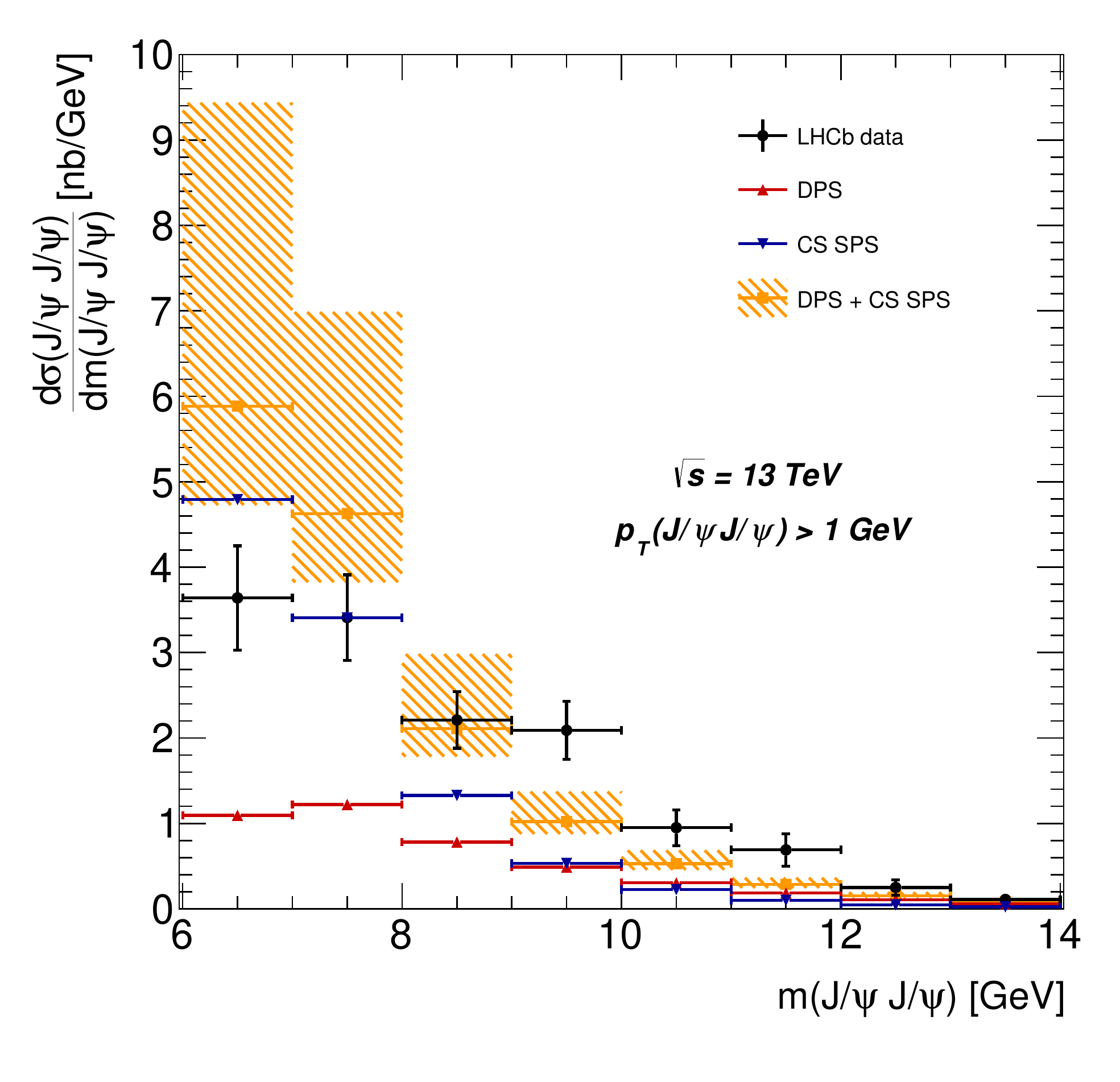}
\caption{Prompt double $J/\psi$ production 
as functions of different kinematical variables 
calculated at $p_T(J/\psi,J/\psi) > 1$~GeV and $\sqrt s = 13$~TeV.
Other kinematical cuts applied are described in the text.
The A0 gluon distribution in proton is used.}
\label{fig_LHCb2}
\end{center}
\end{figure}

\begin{figure}
\begin{center}
\includegraphics[width=6.0cm]{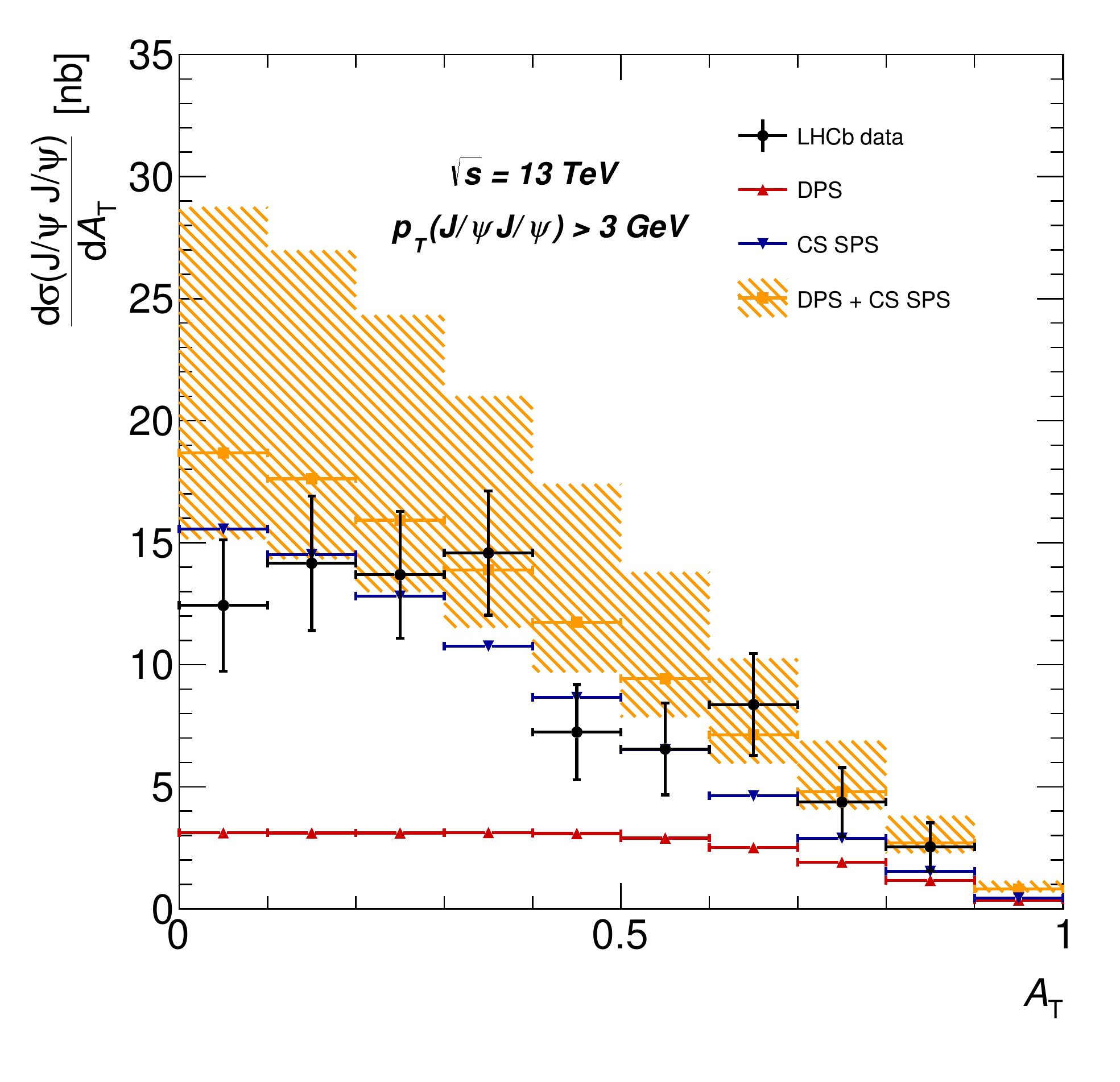}
\includegraphics[width=6.0cm]{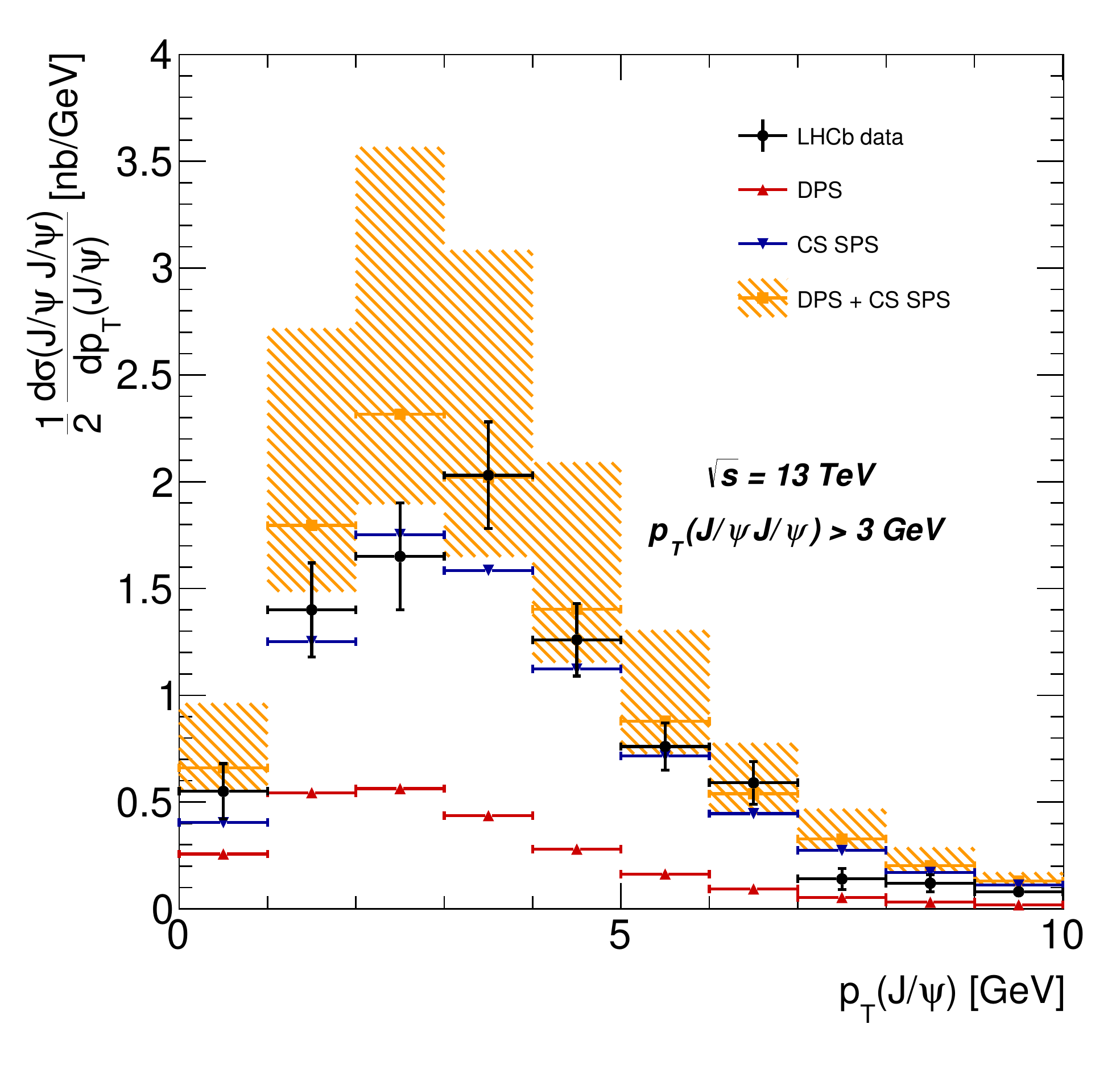}
\includegraphics[width=6.0cm]{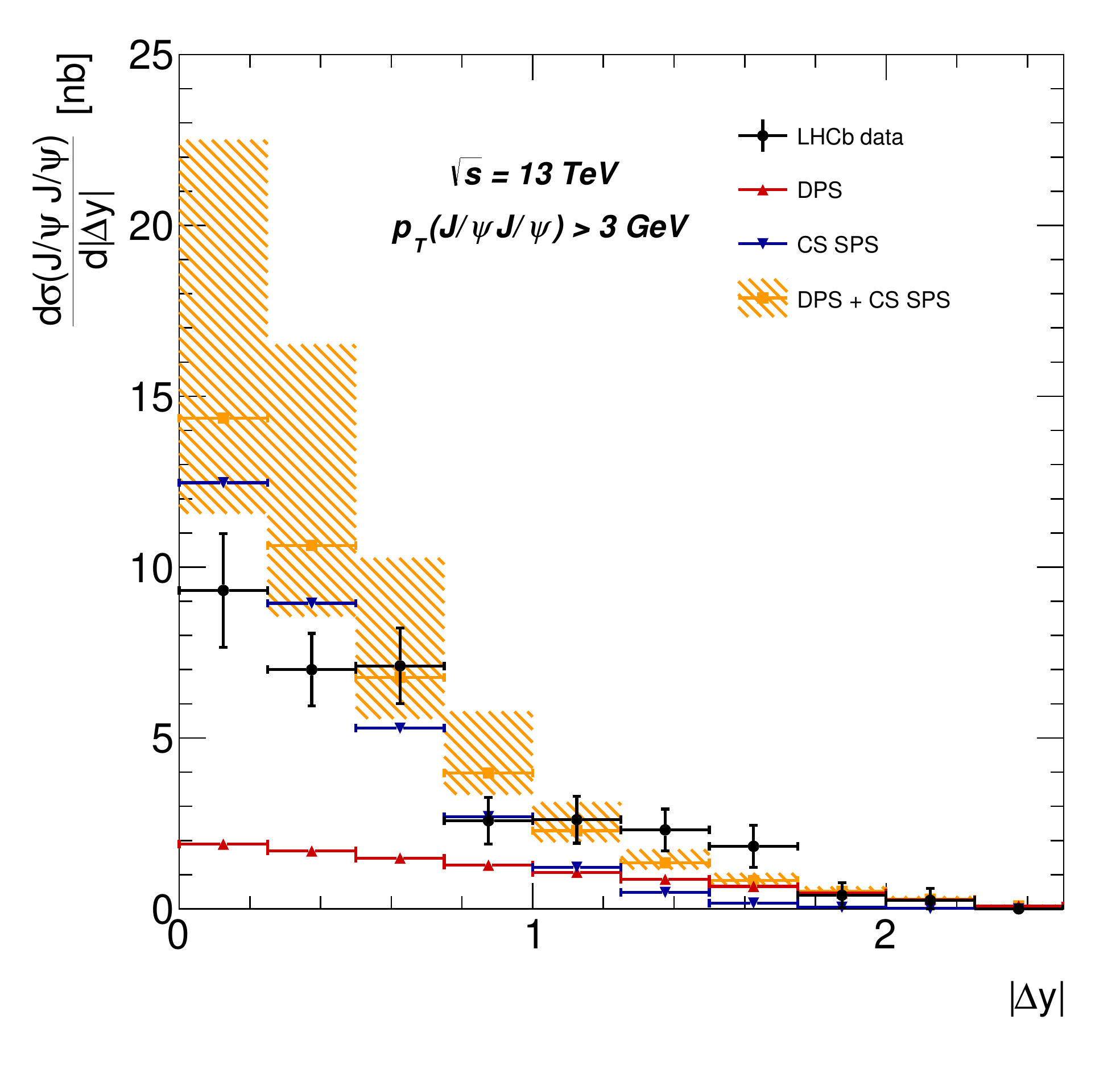}
\includegraphics[width=6.0cm]{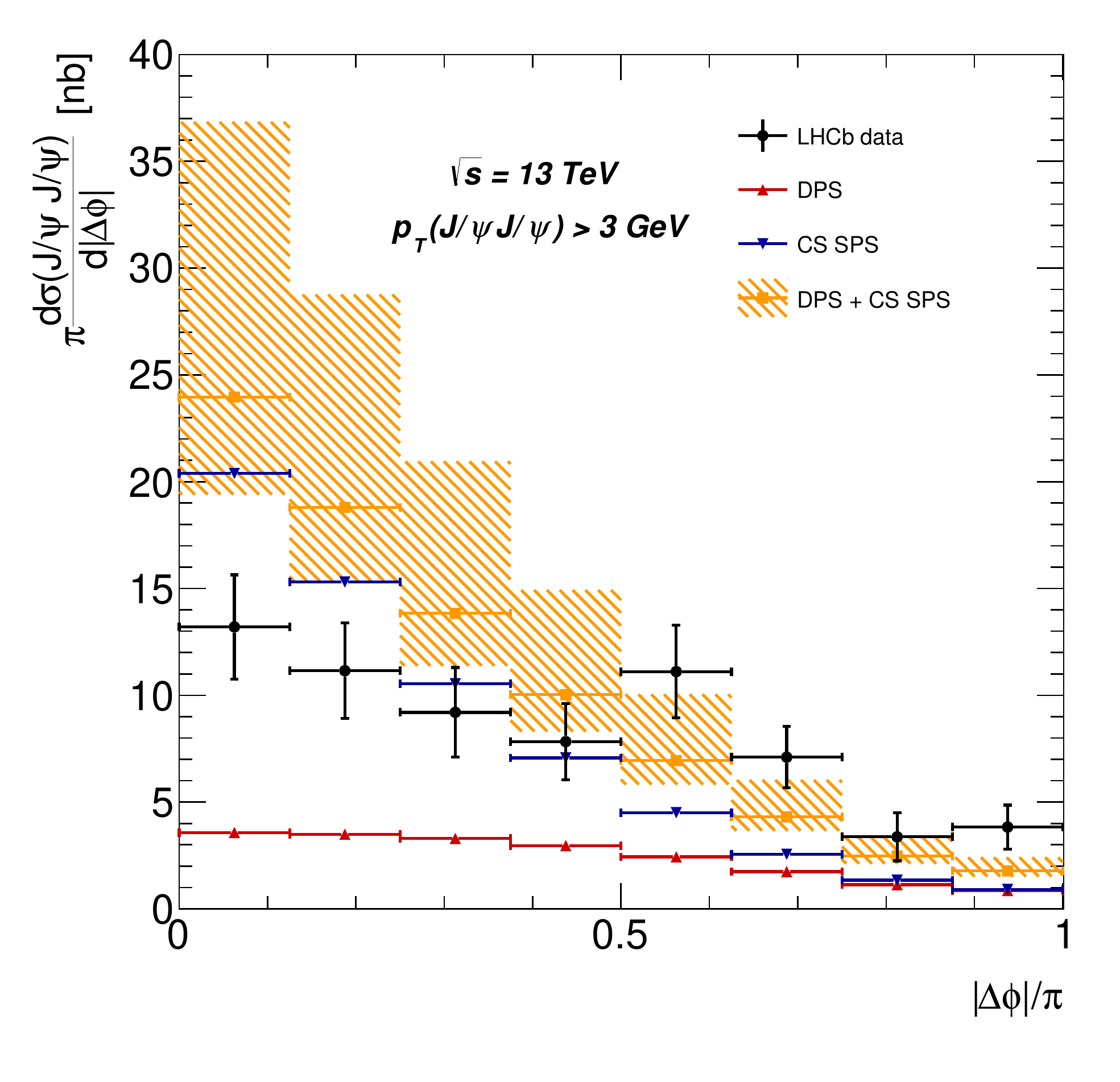}
\includegraphics[width=6.0cm]{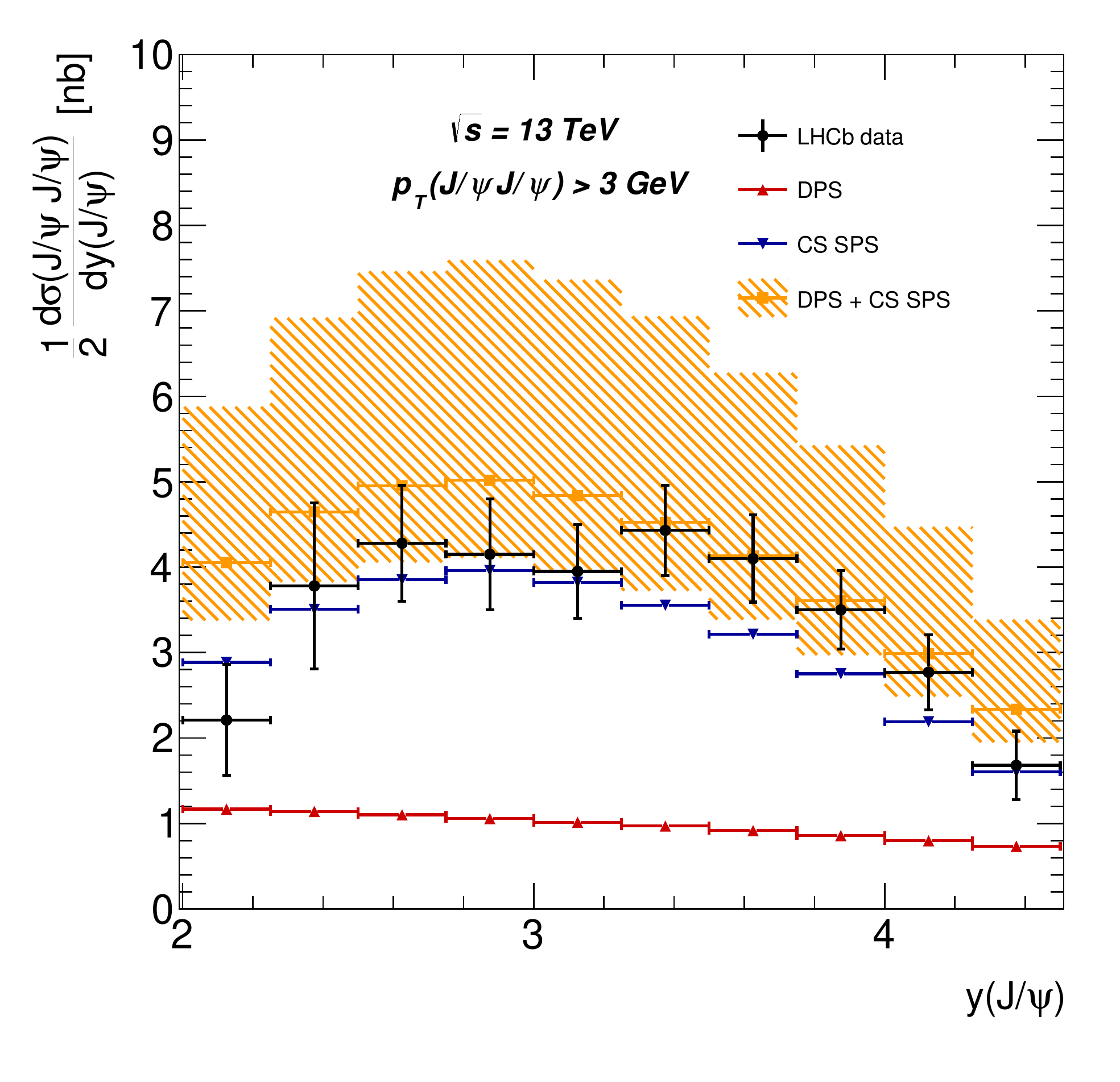}
\includegraphics[width=6.0cm]{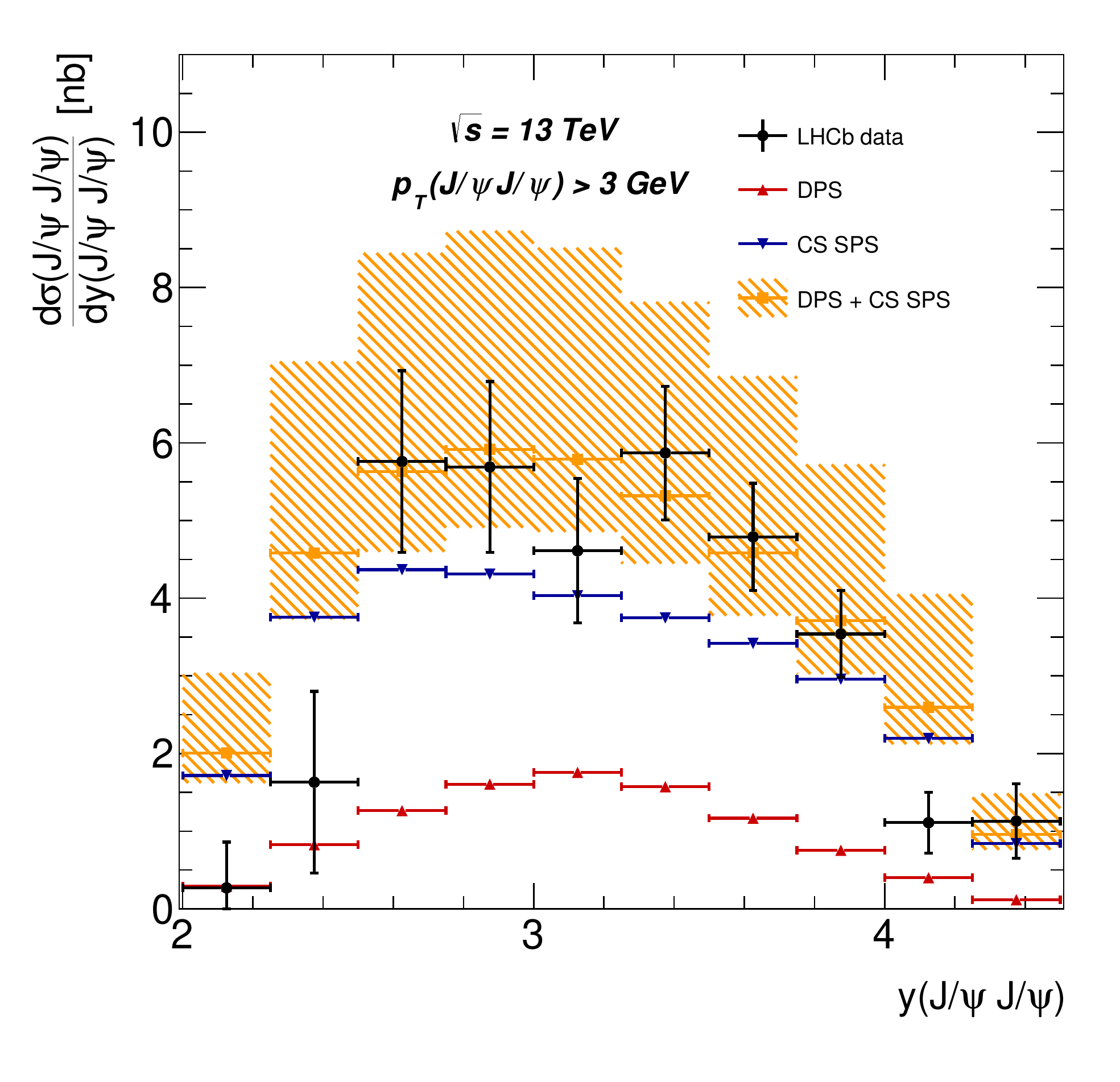}
\includegraphics[width=6.0cm]{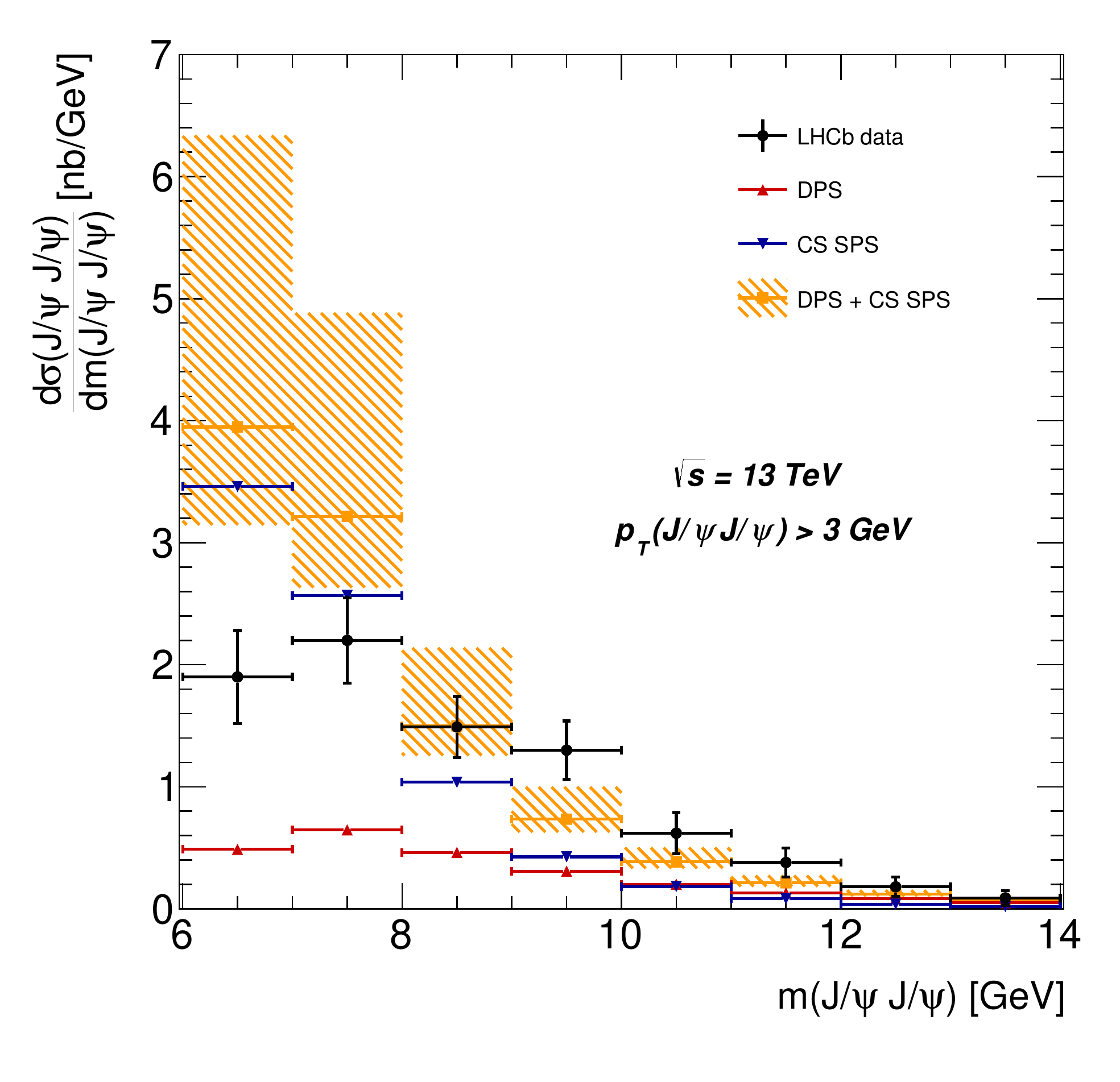}
\caption{Prompt double $J/\psi$ production 
as functions of different kinematical variables 
calculated at $p_T(J/\psi,J/\psi) > 3$~GeV and $\sqrt s = 13$~TeV.
Other kinematical cuts applied are described in the text.
The A0 gluon distribution in proton is used.}
\label{fig_LHCb3}
\end{center}
\end{figure}

A comparison of our predictions with the LHCb experimental data is displayed
in Figs.~\ref{fig_LHCba} --- \ref{fig_LHCb3}. The theoretical uncertainty bands include both scale 
uncertainties and uncertainties coming from the $\sigma_{\rm eff}$ fitting procedure.
First of them have been estimated in a usual way,
by varying the $\mu_R$ scale
around its default value by a factor of $2$.
This was accompanied with using the A0$+$ and A0$-$ gluon densities instead of  
default A0 distribution, in accordance with\cite{59}.
As one can see, we achieved a reasonably good
agreement between the results of our calculations and 
LHCb measurements, both for $\sqrt s = 7$ and $13$~TeV.
There is only exception in the threshold region, $m(J/\psi, J/\psi) \leq 9$~GeV,
where our predictions systematically overshoot the data.
However, at such low $m(J/\psi, J/\psi)$ 
an accurate treatment of
multiple soft gluon emissions, relativistic corrections and other
nonperturbative effects becomes necessary
to produce the theoretical estimations. All these issues are 
out from our present consideration.
Next, we find that neither the SPS terms, nor the DPS contributions alone
are able to describe the LHCb data, but only their sum.
In particular, the DPS contributions are essential to reproduce 
the measured $|\Delta y (J/\psi, J/\psi)|$ distributions 
at $|\Delta y (J/\psi, J/\psi)| \geq 1$ or $1.5$,
that confirms the previous expectations\cite{35,36,37}.
They are important to 
describe also the normalization
of $J/\psi$ rapidity distributions and
shape of transverse momentum asymmetry ${\cal A}_T$ at ${\cal A}_T \leq 0.4$, see Figs.~\ref{fig_LHCb1} --- \ref{fig_LHCb3}.

The presented results, being considered altogether with the ones for inclusive
single production of charmonia states\cite{19},
can give a significant impact on the
understanding of charmonia production 
within the NRQCD framework and, in particular, on the further understanding of DPS mechanism.
The most interesting outcome of our study is that
the extremely low value of DPS effective cross section, $\sigma_{\rm eff} \sim 2- 5$~mb, obtained 
in earlier analyses of double $J/\psi$ production at the LHC, is not confirmed.

\section{Conclusion} \indent

We have considered the prompt production of $J/\psi$ meson pairs in $pp$ collisions at the LHC
using the $k_T$-factorization approach of QCD.
We employ the fragmentation mechanism to evaluate the color octet 
contributions to the production cross sections and
take into account the combinatorial effects of multiple gluon radiation
in the initial state using the CCFM evolution equation.
The latter could be essential
in the kinematical region covered by the CMS and ATLAS measurements.
On the other hand, we have demonstrated that
the experimental data taken by the LHCb Collaboration
at forward rapidities can be described well by the color singlet terms and contributions 
from the double parton scattering mechanism.
We determine the DPS effective cross section 
$\sigma_{\rm eff} = 17.5 \pm 4.1$~mb
from the combined analysis of the LHCb data collected 
at $\sqrt s = 7$ and $13$~TeV.
The extracted value is compatible with many other estimations 
based on essentially different final states.
The extremely low $\sigma_{\rm eff} \sim 2 - 5$~mb, obtained earlier
from the double $J/\psi$ production data, is not confirmed.

\section*{Acknowledgements} \indent

The authors thank G.I.~Lykasov for very useful discussions on the topic. 
M.A.M. and A.A.P. were supported by grants of the foundation for the advancement 
of theoretical physics and mathematics "Basis" 20-1-3-11-1 and 18-2-6-129-1, respectively.

\newpage

\end{document}